\documentclass[aps,11pt]{revtex4}
\usepackage{docs}
\usepackage{epsfig}
\usepackage{amsmath}
\usepackage{amsfonts}
\usepackage{bm}
\usepackage{times}
\usepackage{graphicx}
\usepackage{axodraw}

\def\lsim{\:\raisebox{-0.5ex}{$\stackrel{\textstyle<}{\sim}$}\:}
\def\gsim{\:\raisebox{-0.5ex}{$\stackrel{\textstyle>}{\sim}$}\:}
\def\half{{\textstyle{1 \over 2}}}
\def\fourth{{\textstyle{1 \over 4}}}
\def\bold#1{\setbox0=\hbox{$#1$}%
     \kern-.025em\copy0\kern-\wd0
     \kern.05em\copy0\kern-\wd0
     \kern-.025em\raise.0433em\box0 }

\begin{document}

\begin{flushright}
SHEP-10-33\\
USM-TH-276
\end{flushright}

%\preprint{SHEP-10-33}
%\preprint{USM-TH-276}

%%%%%%%%%%%%%%%%%%%%%%%%%%%%%%%%%%%%%%%%%%%%%%%%%%%%%%%%%%%%%%%%%%%%%%
\title{Fermiophobia in a Higgs Triplet Model}
%%%%%%%%%%%%%%%%%%%%%%%%%%%%%%%%%%%%%%%%%%%%%%%%%%%%%%%%%%%%%%%%%%%%%%
\author{A.G. Akeroyd$^{a,b}$}
\author{ Marco  A.  D\'{\i}az$^c$}
\author{Maximiliano A.   Rivera$^{d}$}
\author{Diego Romero$^c$}

\affiliation{ $^a$  Department of  Physics and Center  for Mathematics
  and  Theoretical  Physics,  National  Central  University,  Chungli,
  Taiwan 320,  Taiwan.\\ $^b$ NExT Institute and School of Physics and 
Astronomy, University of Southampton \\
Highfield, Southampton SO17 1BJ, United Kingdom.\\
$^c$  Departamento de F\'\i  sica, Pontificia
  Universidad  Cat\'{o}lica de Chile,  Santiago 690441,  Chile.\\ $^d$
  Departamento  de F\'\i  sica, Universidad  T\'ecnica  Federico Santa
  Mar\'\i a, Casilla 110-V, Valpara\'\i so, Chile.}  \date{\today}

%%%%%%%%%%%%%%%%%%%%%%%%%%%%%%%%%%%%%%%%%%%%%%%%%%%%%%%%%%%%%%%%%%%%%%
\begin{abstract}

A Fermiophobic Higgs boson can  arise in models with an extended Higgs
sector,   such  as  models   with  scalars   in  an   isospin  triplet
representation.   In  a  specific  model  with a  scalar  triplet  and
spontaneous  violation of lepton  number induced  by a  scalar singlet
field, we show that fermiophobia is not a fine-tuned situation, unlike
in  Two  Higgs  Doublet   Models.  We  study  distinctive  signals  of
fermiophobia which can be probed at  the LHC.  For the case of a small
Higgs   mass   a   characteristic   signal   would   be   a   moderate
$B(H\rightarrow\gamma\gamma)$  accompanied by a  large $B(H\rightarrow
JJ)$ (where  $J$ is a Majoron),  the latter being  an invisible decay.
For  the case  of  a large  Higgs  mass there  is  the possibility  of
dominant $H\rightarrow ZZ, WW$  and suppressed $H\rightarrow JJ$ decay
modes.   In  this  situation,  $B(H\rightarrow  ZZ)$  is  larger  than
$B(H\rightarrow WW)$, which differs from the SM prediction.

\end{abstract}

\maketitle
%%%%%%%%%%%%%%%%%%%%%%%%%%%%%%%%%%%%%%%%%%%%%%%%%%%%%%%%%%%%%%%%%%%%%%
\section{Introduction}
%%%%%%%%%%%%%%%%%%%%%%%%%%%%%%%%%%%%%%%%%%%%%%%%%%%%%%%%%%%%%%%%%%%%%%
%
The Standard Model (SM) of the electroweak and strong interactions is
a very successful  model, although the Higgs sector  still needs to be
probed  by  experiments.   The  LEP  lower bound  on  the  Higgs  mass
$m_H>114.4$ GeV \cite{Barate:2003sz} has been complemented at Fermilab
by ruling  out the region between  160 and 170  GeV \cite{:2009pt}. In
the  meantime  the  Large  Hadron  Collider  (LHC)  experiments  ATLAS
\cite{Aad:2009wy} and  CMS \cite{:2008zzk}  will soon join  the search
for the  Higgs boson. In the SM,  the Higgs sector is  composed of one
Higgs doublet  under $SU(2)_L$, nevertheless,  there is no  reason why
the  Higgs sector may  not be  larger, and  extensions are  very often
explored  \cite{Accomando:2006ga}.  Higgs  bosons  in isospin  triplet
representations  \cite{Georgi:1985nv}  have   been  studied,  and  are
primarily   motivated  by   a  neutrino   mass   generation  mechanism
\cite{Konetschny:1977bn},  for example  via  spontaneous violation  of
lepton number \cite{Schechter:1981cv}.  The phenomenology of the model
has been  well studied \cite{Diaz:1998zg}, and  more recently, renewed
attention has been given to  the detection prospects of the doubly and
singly  charged scalars  at  the LHC  \cite{Chun:2003ej} (for  earlier
studies see e.g. \cite{Gunion:1989in}).

Fermiophobic  Higgs bosons  \cite{Weiler:1987an}, {\sl  i.e.}, neutral
Higgs  bosons with  negligible  couplings to  fermions,  can arise  in
models with  Higgs triplets,  as well as  in two Higgs  doublet models
(2HDM).   A  Higgs  boson  of  this  type,  denoted  by  $h_f$,  would
dominantly  decay via  $h_f\rightarrow\gamma\gamma$ for  $m_f\lsim 95$
GeV,  and  via $h_f\rightarrow  W^+W^-$  and  $h_f\rightarrow ZZ$  for
$m_f\gsim 95$ GeV, for a Higgs  with SM like couplings to gauge bosons
\cite{Stange:1994ya}.   In the  latter case,  the decay  rates satisfy
$\Gamma(h_f\rightarrow  W^+W^-)/\Gamma(h_f\rightarrow  ZZ)\gsim2$,  in
the region of large Higgs mass \cite{Carena:2002es}.

In this article  we study the appearance of  fermiophobic Higgs bosons
in a particular  Higgs Triplet Model (HTM) that  includes a singlet, a
doublet  and a  triplet  Higgs field,  the  so-called ``123  models''.
These models are characterized by  a spontaneous violation of a global
$U(1)$ symmetry  through a vacuum expectation value  of a $SU(2)\times
U(1)$ Higgs singlet $\langle  \sigma \rangle$.  Therefore, this broken
symmetry  produces a  massless Goldstone  Boson called  a  Majoron (J).
Within  this model  we  show  that fermiophobia  is  not a  fine-tuned
situation as  in the 2HDM. In  fact, the model has  a tendency towards
fermiophobia  mainly  due  to   the  hierarchy  of  the  three  vacuum
expectation values. Furthermore, we  emphasize a scenario in which the
decay of  the fermiophobic Higgs boson  to Majorons via $h  \to JJ$ is
partially   suppressed,  thereby  allowing   branching  ratios   of  a
fermiophobic Higgs into  gauge bosons which can be  probed by the LHC.
Our  work is organized  as follows.  In section  \ref{sec:HTM} we
  introduce the  Higgs Triplet Model  and in Section  \ref{sec:FM} the
  scenario of fermiophobia with  Majoron suppression is described.  In
  section \ref{sec:fdecay} the decays  of the fermiophobic Higgs boson
  are    discussed,   with    phenomenology    studied   in    section
  \ref{sec:pheno}.     Conclusions    are    contained   in    section
  \ref{sec:concl}.

%
%%%%%%%%%%%%%%%%%%%%%%%%%%%%%%%%%%%%%%%%%%%%%%%%%%%%%%%%%%%%%%%%%%%%%%
\section{Higgs Triplet Model}
\label{sec:HTM}
%%%%%%%%%%%%%%%%%%%%%%%%%%%%%%%%%%%%%%%%%%%%%%%%%%%%%%%%%%%%%%%%%%%%%%
% 
The Higgs Triplet Model  (HTM) which we will study \cite{Diaz:1998zg},
includes  a complex triplet  Higgs field  $\Delta$ with  lepton number
$L=-2$ and  hypercharge $Y=2$, a  complex doublet Higgs  field $\phi$,
with lepton number $L=0$ and hypercharge $Y=-1$,
\begin{equation}
\Delta                    =  \left[\begin{array}{cc}
    \Delta^0&\Delta^+/\sqrt2\\
\Delta^+/\sqrt2&\Delta^{++}\end{array}\right],
\qquad\qquad\phi=\left[\begin{array}{c}\phi^0\\
 \phi^-\end{array}\right],
\label{higgs}
\end{equation}
and a complex  singlet Higgs field $\sigma$, with  lepton number $L=2$
and  hypercharge  $Y=0$.  The  model  without  the  singlet field  has
received much  attention recently \cite{Chun:2003ej}, and  we note the
phenomenology of the charged scalars (doubly and singly) at the LHC is
essentially identical in both models.
%
%%%%%%%%%%%%%%%%%%%%%%%%%%%%%%%%%%%%%%%%%%%%%%%%%%%%%%%%%%%%%%%%%%%%%%
\subsection{Higgs Potential and Mass Spectrum}
%%%%%%%%%%%%%%%%%%%%%%%%%%%%%%%%%%%%%%%%%%%%%%%%%%%%%%%%%%%%%%%%%%%%%%
%
The scalar potential can be written as follows:
\begin{eqnarray}
V(\phi,\Delta,\sigma)&=&
\mu_1^2\sigma^\dagger\sigma+\mu_2^2\phi^\dagger\phi+\mu_3^2\,
{\rm tr}(\Delta^\dagger\Delta)+
\lambda_1(\phi^\dagger\phi)^2+\lambda_2[{\rm tr}
(\Delta^\dagger\Delta)]^2
\nonumber\\
&&+\lambda_3\phi^\dagger\phi\,{\rm tr}(\Delta^\dagger\Delta)+
\lambda_4\,{\rm tr}(\Delta^\dagger\Delta\Delta^\dagger\Delta)+
\lambda_5(\phi^\dagger\Delta^\dagger\Delta\phi)\nonumber\\
&&+\beta_1(\sigma^\dagger\sigma)^2+
\beta_2(\phi^\dagger\phi)(\sigma^\dagger\sigma)
+\beta_3\,{\rm tr}(\Delta^\dagger\Delta)\sigma^\dagger\sigma \nonumber\\
&&-
\kappa(\phi^T\Delta\phi\sigma+{\rm h.c.})
\label{eq:GenPot}.
\end{eqnarray}\\
where $\mu_i^2$, $i=1,2,3$,  are mass squared parameters, $\lambda_i$,
$i=1,...,5$ are  dimensionless couplings  not related to  the singlet,
$\beta_i$,  $i=1,2,3$  are  dimensionless  couplings  related  to  the
singlet, and $\kappa$ is a dimensionless coupling that mixes all three
Higgs fields.

The  electroweak symmetry  is  spontaneously broken  when the  neutral
components  of  the Higgs  fields  acquire  vacuum expectation  values
$v_i$, $i=1,2,3$. We shift the Higgs fields in the following way,
\begin{eqnarray}
\sigma&=&\frac{v_1}{\sqrt2}+\frac{R_1+iI_1}{\sqrt{2}}\nonumber\\
\phi^0&=&\frac{v_2}{\sqrt{2}}+\frac{R_2+iI_2}{\sqrt{2}}
\label{shift}\\
\Delta^0&=&\frac{v_3}{\sqrt2}+\frac{R_3+iI_3}{\sqrt{2}}\nonumber
\end{eqnarray}
finding  minimization  conditions,  or tree-level  tadpole  equations,
given by,
\begin{equation}
V_{lineal}=t_1R_1+t_2R_2+t_3R_3=0\,,
\label{eq:Vlinear}
\end{equation}
where
\begin{eqnarray}
t_1&=&v_1(\mu_1^2+\beta_1 v_1^2 + \half \beta_2 v_2^2 
+\half\beta_3 v_3^2)-
\half\kappa v_3v_2^2\nonumber\\
t_2&=&v_2(\mu_2^2+\lambda_1 v_2^2+\half \lambda_3 v_3^2 
+\half\lambda_5 v_3^2+
\half\beta_2v_1^2-\half\kappa v_1v_3)
\label{eq:tadpoles}\\
t_3&=&v_3(\mu_3^2+\lambda_2v_3^2+\half\lambda_3v_2^2+\lambda_4 v_3^2+
\half\lambda_5
v_2^2+\half\beta_3v_1^2)-\half\kappa v_1v_2^2
\nonumber.
\end{eqnarray}
In  ref.~\cite{Diaz:1998zg} cases  where different  vacuum expectation
values are  equal to  zero are analyzed,  but these scenarios  are not
relevant for  our purposes. In the  following we assume  all vev's are
non zero.

The quadratic potential can be written as follows,
\begin{eqnarray}
V_{quadratic}=\half\Big[R_1,R_2,R_3\Big]{\bold M^2_R}
\left[\begin{matrix}
R_1 \cr R_2 \cr R_3
\end{matrix}\right]
+\half\Big[I_1,I_2,I_3\Big]
{\bold M^2_I}\left[\begin{matrix}
I_1 \cr I_2 \cr I_3
\end{matrix}\right]+\nonumber\\
\Big[\phi^-,\Delta^-\Big]{\bold M^2_+}
\left[\begin{matrix}
\phi^+ \cr \Delta^+
\end{matrix}\right]
+m^2_{\Delta^{++}}\Delta^{++}\Delta^{--}
\label {eq:NeutScalLag}
\end{eqnarray}
The CP-even neutral Higgs mass matrix is given by,
\begin{equation}
{\bold M_R^2}=\left[
\begin{array}{ccc}
2\beta_1v_1^2+\half\kappa v_2^2\frac{v_3}{v_1}+\frac{t_1}{v_1}&
\beta_2v_1v_2-\kappa v_2v_3 
&\beta_{3}v_1v_3-\half\kappa v_2^2\\
\beta_2v_1v_2-\kappa v_2v_3&2\lambda_1v_2^2+\frac{t_2}{v_2}
&(\lambda_3+\lambda_5)v_2v_3-\kappa v_1v_2\\
\beta_{3}v_1v_3-\half\kappa v_2^2&(\lambda_3+\lambda_5)v_2v_3-
\kappa v_1v_2
&2(\lambda_2+\lambda_4)v_3^2+\half\kappa v_2^2\frac{v_1}{v_3}+
\frac{t_3}{v_3}\\
\end{array}  \right]
\label{MR}
\end{equation}
where  we have  eliminated  the mass  parameters  $\mu_i^2$ using  the
tadpole equations.  This mass matrix is diagonalized  by an orthogonal
matrix $O_R$, which can be parametrized with three angles,
\begin{eqnarray}
O_R &=&
\left[\begin{matrix}
1 & 0 & 0 \cr 0 & c_{23} & s_{23} \cr 0 & -s_{23} & c_{23}
\end{matrix}\right]
\left[\begin{matrix}
c_{13} & 0 & s_{13} \cr 0 & 1 & 0 \cr -s_{13} & 0 & c_{13}
\end{matrix}\right]
\left[\begin{matrix}
c_{12} & s_{12} & 0 \cr -s_{12} & c_{12} & 0 \cr 0 & 0 & 1
\end{matrix}\right] \nonumber \\ 
&=&
\left[\begin{matrix}
c_{13}c_{12} & c_{13}s_{12} & s_{13} \cr 
-c_{23}s_{12}-s_{23}s_{13}c_{12}&c_{23}c_{12}-s_{23}s_{13}s_{12}
&s_{23}c_{13}\cr 
s_{23}s_{12}-c_{23}s_{13}c_{12}&-s_{23}c_{12}-c_{23}s_{13}s_{12}
&c_{23}c_{13}
\end{matrix}\right] \label{Oangles}
\end{eqnarray}
where    $s_{12}=\sin\theta_{12}$,    $c_{12}=\cos\theta_{12}$,    and
similarly for the other two angles $\theta_{13}$ and $\theta_{23}$.

The CP-odd neutral Higgs mass matrix is,
\begin{equation}
{\bold M_I^2}=\left[\begin{array}{ccc}
\half\kappa v_2^2\frac{v_3}{v_1}+\frac{t_1}{v_1} & \kappa v_2v_3 & 
\half\kappa v_2^2\\
\kappa v_2v_3 & 2\kappa v_1v_3+\frac{t_2}{v_2} & \kappa v_1v_2 \\
\half\kappa v_2^2 & \kappa v_1v_2 & \half\kappa 
v_2^2\frac{v_1}{v_3}+\frac{t_3}{v_3}
\end{array}\right].
\label{eq:MassMatrixI}
\end{equation}
Clearly, this  mass matrix has two  zero eigenvalues, of  which one is
unphysical and  corresponds to the neutral Goldstone  boson. The other
one  is  physical and  corresponds  to  the  Majoron $J$.   The  third
eigenvalue is the CP-odd neutral Higgs $A$, and has a mass given by,
\begin{equation}
m_A^2=\frac{\kappa}{2}\left(\frac{v_1v_2^2}{v_3}+\frac{v_2^2v_3}{v_1}+
4v_1v_3\right)
\label{ma}
\end{equation}
As one can see, a value  for $\kappa\ne0$ is essential in our model in
order to have a massive CP-odd Higgs boson.

The charged Higgs mass matrix, given by
\begin{equation}
{\bold M_+^2}=\half(\kappa v_1v_2-\half\lambda_5v_2v_3)
\left[\begin{array}{cc}
2v_3/v_2 & -\sqrt{2} \\ -\sqrt{2} & v_2/v_3 \\
\end{array}\right].
\label{eq:MassMatrixIII}
\end{equation}
also  has a zero  eigenvalue, corresponding  to the  charged Goldstone
boson.  It  is diagonalized  by an orthogonal  matrix $O_+$  such that
$O_+\,{\bold M_+^2}\,O_+^T={\mathrm{diag}}(m_{H^+}^2,0)$. The rotation
matrix is,
\begin{equation}
O_+=\left[\begin{matrix} c_+ & s_+ \cr -s_+ & c_+ \end{matrix}\right]
=\frac{1}{\sqrt{v_2^2+2v_3^2}}\left[\begin{matrix}
\sqrt{2}v_3 & -v_2 \cr v_2 & \sqrt{2}v_3 \end{matrix}\right]
\end{equation}
where  $s_+=\sin\theta_+$, $c_+=\cos\theta_+$,  and $\theta_+$  is the
angle of rotation. The massive  eigenvalue is the singly charged Higgs
boson, with a mass,
\begin{equation}
m_{H^+}^2=\frac{1}{2}\left(\kappa \frac{v_1}{v_3}
-\frac{1}{2}\lambda_5\right)\left(v_2^2+2v_3^2\right)
\end{equation}
Finally, the doubly charged Higgs boson has the following mass,
\begin{equation}
m_{\Delta^{++}}^2=\half\kappa\frac{v_1v_2^2}{v_3}-\half\lambda_5v_2^2
-\lambda_4v_3^2
\label{DeltaMass}
\end{equation}
%

%
%%%%%%%%%%%%%%%%%%%%%%%%%%%%%%%%%%%%%%%%%%%%%%%%%%%%%%%%%%%%%%%%%%%%%%
\subsection{Gauge Sector}
%%%%%%%%%%%%%%%%%%%%%%%%%%%%%%%%%%%%%%%%%%%%%%%%%%%%%%%%%%%%%%%%%%%%%%
%
The kinetic terms of the Higgs fields are,
\begin{equation}
{\cal L}_{\mathrm{kinetic}}=(D_\mu\phi)^\dagger(D^\mu\phi)+
{\mathrm{Tr}}\left[(D_\mu\Delta)^\dagger(D^\mu\Delta)\right]+
\partial_\mu\sigma^\dagger \partial^\mu\sigma
\end{equation}
where the covariant derivatives can be written as,
\begin{equation}
D_\mu=\partial_\mu+ig{\bf{T}}_a W_\mu^a+i\half g'{\bf{Y}}B_\mu
\end{equation}
where the  action of the isospin and  hypercharge operators ${\bf{T}}$
and ${\bf{Y}}$ on the Higgs doublet and triplet is
\begin{eqnarray}
{\bf{T}}_a\phi \,\,\,=& \half \tau_a \phi \,,\qquad
{\bf{T}}_a\Delta &=\,\,\, -\half\tau_a^*\Delta -\half\Delta\tau_a
\nonumber\\
{\bf{Y}}\phi \,\,\,=& \!\!\!-\phi \,,\qquad\quad
{\bf{Y}}\Delta &=\,\,\, 2\Delta
\end{eqnarray}
Gauge  bosons  receive contributions  to  their  masses  from the  the
doublet  and  triplet.  After   these  scalar  fields  acquire  vacuum
expectation values, we find,
\begin{equation}
m_W^2=\frac{1}{4}g^2(v_2^2+2v_3^2)\,,\qquad
m_Z^2=\frac{1}{4}(g^2+g'^2)(v_2^2+4v_3^2)
\label{gaugeMasses}
\end{equation}
which leads to the following $\rho$-parameter at tree-level,
\begin{equation}
\rho=1-\frac{2v_3^2}{v_2^2+4v_3^2}
\label{rho}
\end{equation}
The   experimental  measurement   of   $\rho$  is   given  by   $\rho=
1.0002{{+0.0007}\atop{-0.0004}}$, and this  restricts the value of the
triplet vev to  be smaller than a few GeV.   Nevertheless, in order to
satisfy  stringent  bounds  from   astrophysics,  we  will  work  with
$v_3<0.35$  GeV \cite{Diaz:1998zg}.  From  eq.~(\ref{ma}) we  see that
the  small value  for $v_3$  implies  in turn  a small  value for  the
coupling $\kappa$ in  order to have a CP-odd Higgs  mass $m_A$ below 1
TeV.  Another consequence of  the small  value for  $v_3$ is  that the
vacuum expectation value of the Higgs doublet $v_2$ will be very close
to   246  GeV,   as  indicated   by   the  gauge   bosons  masses   in
eq.~(\ref{gaugeMasses}).
%
%%%%%%%%%%%%%%%%%%%%%%%%%%%%%%%%%%%%%%%%%%%%%%%%%%%%%%%%%%%%%%%%%%%%%%
\section{Fermiophobia and Majoron Suppression}
\label{sec:FM}
%%%%%%%%%%%%%%%%%%%%%%%%%%%%%%%%%%%%%%%%%%%%%%%%%%%%%%%%%%%%%%%%%%%%%%
%
Fermiophobia  was first introduced  in \cite{Weiler:1987an},  where it
was stressed that  the mechanism for the generation  of fermion masses
could be independent of the mechanism for the generation of the masses
of the  gauge bosons (e.g. a  2HDM in which the  fermions receive mass
from just one vacuum expectation value, while the gauge bosons receive
mass from both).

We investigate  the possibility that the lightest  CP-even Higgs boson
is  fermiophobic,   in  which  case  the   conventional  decay  modes,
$H_1^0\rightarrow  b\overline{b}, \tau^-\tau^+$,  are  suppressed.  In
addition, we  study the singlet  content of this Higgs  boson, looking
for cases in which the fermiophobic Higgs has a suppressed mixing with
the singlet field, a  situation which we call ``Majoron suppression''.
If the fermiophobic Higgs has a large mixing with the singlet, it will
decay  mainly into  two  Majorons,  which leads  to  a missing  energy
signature.  In this paper we focus on signatures of fermiophobia which
are visible in detectors.

We start by doing a general scan of the parameter space as follows:
\begin{eqnarray}
0.5\,{\mathrm{GeV}}<v_1<1\,{\mathrm{TeV}}\,,\qquad & 
v_2=246\,{\mathrm{GeV}}\,, &
\qquad v_3<0.35\,{\mathrm{GeV}}\,,\nonumber\\
0<\kappa<0.1\,,\qquad & |\beta_{1,2,3}|<4\,, & \qquad 
|\lambda_{1...5}|<4\,.
\label{scanDef}
\end{eqnarray}
Our aim  is to see how  large the parameter space  is for fermiophobia
and  Majoron suppression  for the  lightest CP-even  Higgs  boson.  In
addition to  eq.~(\ref{scanDef}), we respect  the current experimental
limits for  the masses of  $A^0$, $H^\pm$, and  {\bf $\Delta^{\pm\pm}$
}~($m_A>90$ GeV,~ $m_{H^+}>80$ GeV,~ $m_{\Delta^{++}}>136$ GeV), and
we also require $90<m_{H_1}<300$ GeV.

%------------------------ FIGURE -------------------------------------
\begin{figure}[h!]
\centerline{
\protect
\hbox{
\epsfig{file=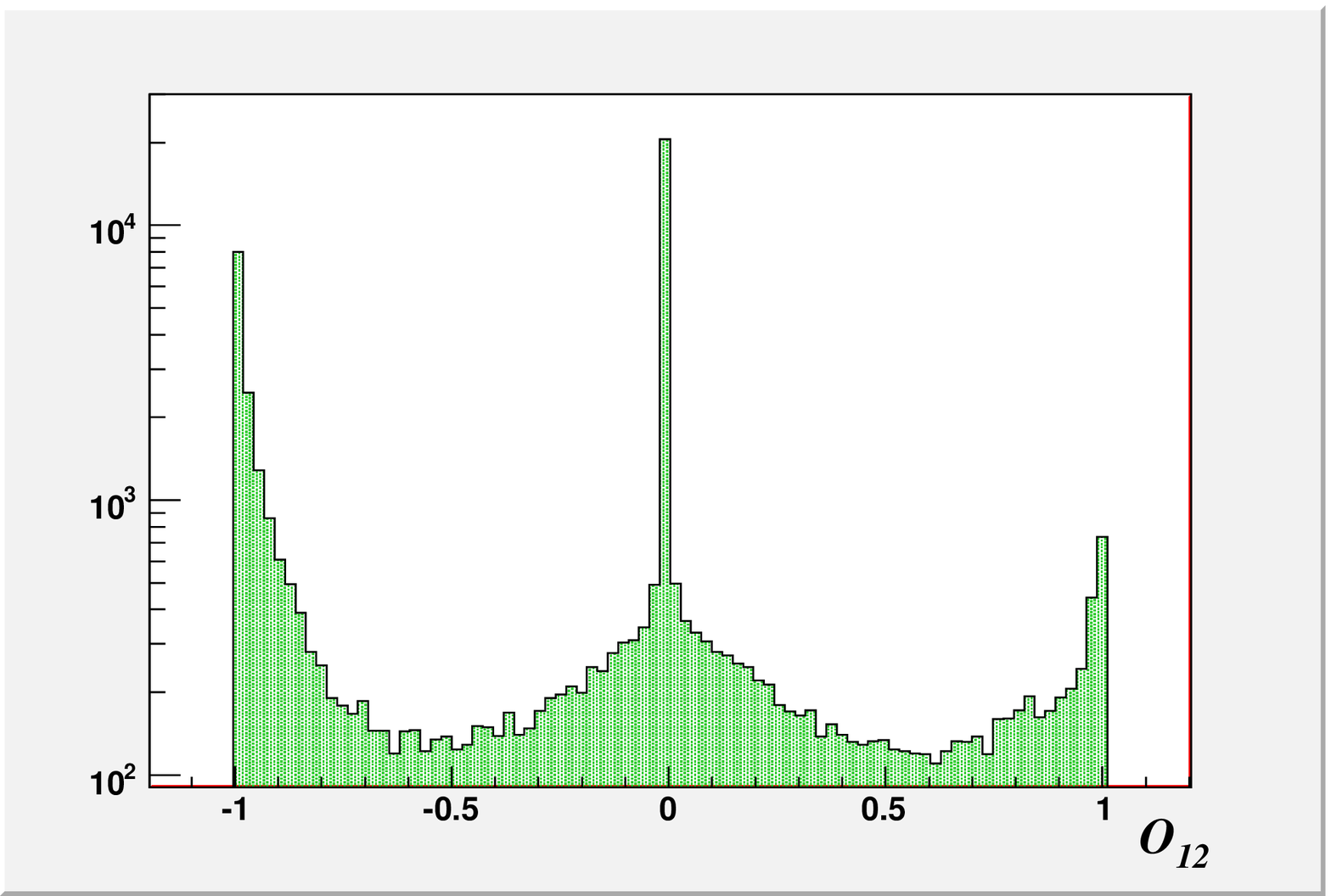,width=0.50\textwidth}
\epsfig{file=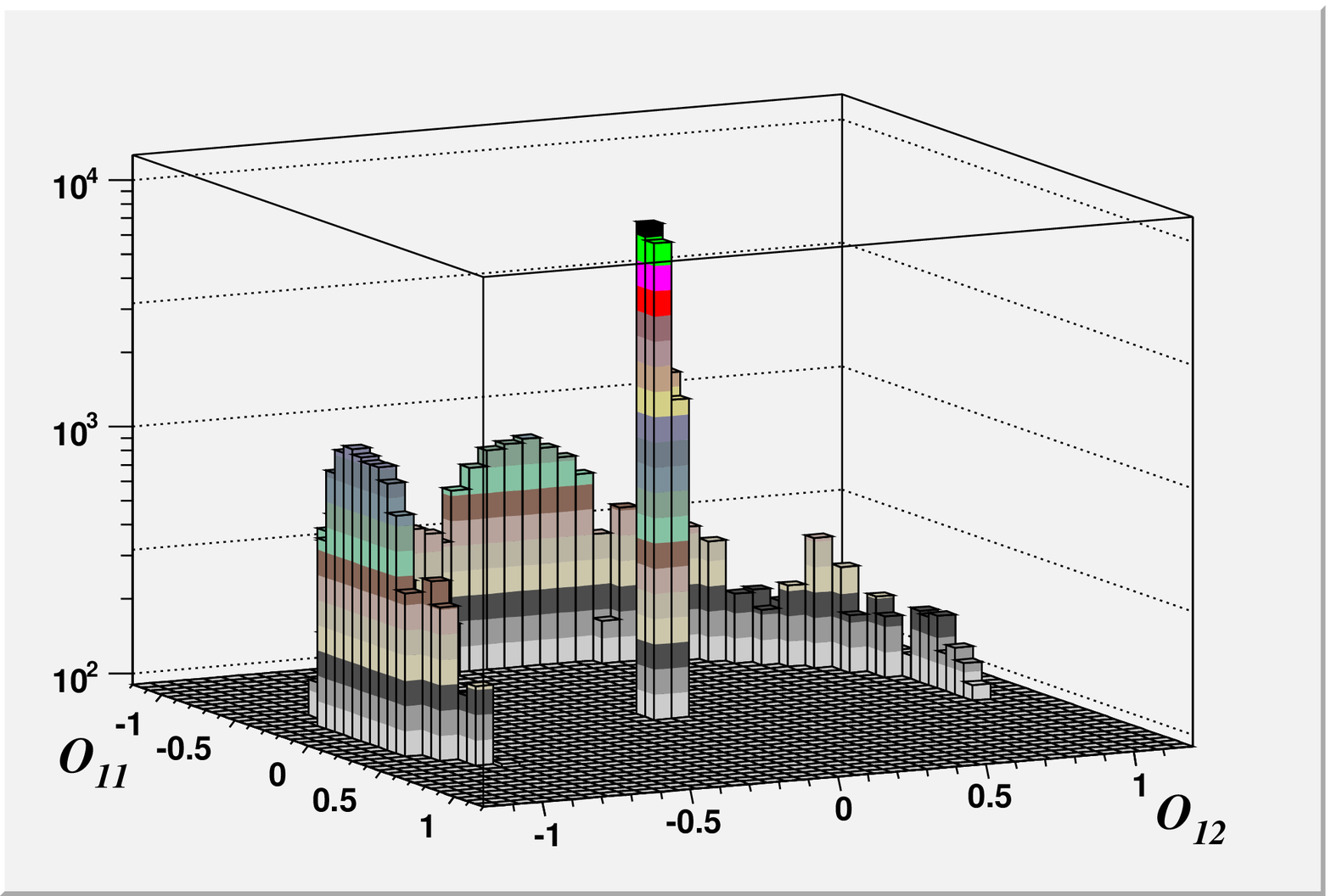,width=0.50\textwidth}}}
\caption{\it  Frequency  histogram  for  $O_R^{12}$ (left  frame)  and
  $O_R^{11}$-$O_R^{12}$ (right  frame) in a general  scan of parameter
  space.}
\label{O1andO2}
\end{figure}
%---------------------------------------------------------------------
%

In the left frame of  Fig.~\ref{O1andO2} we show a frequency histogram
from which  one can  see the values  of the matrix  element $O_R^{12}$
(see eq.~(\ref{Oangles})), with no other restriction on the parameters
in the  Lagrangian, except  for a correct  minimization of  the scalar
potential.  In  this plot we see  the rather unexpected  result that a
sharp  maximum   is  obtained  at  $O_R^{12}=0$,   indicating  a  high
concentration  of points  in  the fermiophobic  region.  Therefore  we
conclude that fermiophobia is not a fine-tuned scenario in this model.

In Table \ref{tab:O12} we show the fraction of points from the general
scan of the parameter  space (defined by eq.~(\ref{scanDef})) that lie
in a given region  around the exact fermiophobic point $|O_R^{12}|=0$.
Clearly it is not necessary to deviate too much from $|O_R^{12}|=0$ in
order to  encapsulate an important  number of the points  in parameter
space.   In   other  words,  the   model  has  a   ``preference''  for
fermiophobia. The  reason for  this is that  the CP-even  neutral mass
matrix in  eq.~(\ref{MR}) has diagonal elements which  are much larger
than the non-diagonal  elements in a large region  of parameter space,
making  $O^{12}_R$ naturally small.   This feature  is present  in our
model due to both the hierarchy of the three vacuum expectation values
and the smallness of the $\kappa$ coupling.

%
%------------------------ TABLE  -------------------------------------
\begin{table}[h!]
\begin{center}
\begin{tabular}{c|c}
\hline
Fraction of Scan & Max.~$|O_R^{12}|$ value \\
\hline \hline
45\% & 0,070   \\
40\% & 0,0086  \\
30\% & 0,0014  \\
20\% & 0,00056 \\
10\% & 0,00019 \\
\hline \hline 
\end{tabular}
\caption{Fraction of points in the general scan of the parameter space
  that  are  within   a  given  region  around  the   point  of  exact
  fermiophobia defined by $|O_R^{12}|=0$.} \label{tab:O12}
\end{center}
\end{table}
%---------------------------------------------------------------------
%

In  the right frame  of Fig.~\ref{O1andO2}  we show  a two-dimensional
frequency histogram in the  plane $O_R^{11}-O_R^{12}$, within the same
scan as before. The peak around $O_R^{11}=O_R^{12}=0$ (or equivalently
$O_R^{13}=1$)  corresponds  to a  fermiophobic  Higgs with  suppressed
couplings to  the Majoron,  implying that the  visible decay  modes of
this Higgs  boson are not  suppressed. The concentration  of parameter
space  points around  $O_R^{13}=1$ is  again due  to the  hierarchy of
vacuum expectation values.

%------------------------ TABLE  -------------------------------------
\begin{table}[h!]
\begin{center}
\begin{tabular}{c|c}
\hline
Fraction of Scan & Min.~$|O_R^{13}|$ value \\
\hline \hline
41.5\% & 0.9  \\
41.0\% & 0.999 \\
36.6\% & 0.99999  \\
27.8\% & 0.999999  \\
14.0\% & 0.9999999  \\
\hline \hline 
\end{tabular}
\caption{Fraction of points in the general scan of the parameter space
  that  are  within   a  given  region  around  the   point  of  exact
  fermiophobia     and      Majoron     suppression     defined     by
  $|O_R^{13}|=1$.}\label{tab:O13}
\end{center}
\end{table}
%---------------------------------------------------------------------
%

In  Table \ref{tab:O13}  we display  the fraction  of points  within a
given  region around  the  exact fermiophobic  and Majoron  suppressed
case, defined by $|O_R^{13}|=1$. It is surprising how little deviation
from this  point is  necessary to  find a large  fraction of  the scan
points  around  the  fermiophobic  and Majoron  suppressed  situation,
indicating that it is not a fine-tuned case in this model.
%
%%%%%%%%%%%%%%%%%%%%%%%%%%%%%%%%%%%%%%%%%%%%%%%%%%%%%%%%%%%%%%%%%%%%%%
\subsection{Imposing Fermiophobia in the CP-even Higgs Sector}
%%%%%%%%%%%%%%%%%%%%%%%%%%%%%%%%%%%%%%%%%%%%%%%%%%%%%%%%%%%%%%%%%%%%%%
%
We are interested in the possibility of having a light CP-even neutral
Higgs boson  with suppressed couplings to  fermions (fermiophobia).  A
light    fermiophobic     Higgs    boson    is     characterized    by
$O_R^{12}=c_{13}s_{12}=0$,  since it  is
mainly the Higgs doublet $\phi$ which couples to fermions (the triplet
coupling  to   the  fermions  is  suppressed).    With  the  condition
$s_{12}=0$ we  find general fermiophobia. The  diagonalizing matrix in
this case is,
\begin{eqnarray}
O_R = \left[\begin{matrix}
\pm c_{13}        & 0         & s_{13}        \cr 
\mp s_{23}\,s_{13} & \pm c_{23} & s_{23}\,c_{13} \cr
\mp c_{23}\,s_{13} & \mp s_{23} & c_{23}\,c_{13}
\end{matrix}\right]
\end{eqnarray}
where $\pm$  corresponds to $\mathrm{sign}(c_{12})$.  The diagonalized
CP-even Higgs  mass matrix is  given by $({\bf M}^2_{R})_{diag}  = O_R
{\bf   M}^2_{R}  O_R^T$,   and  implies   the   following  consistency
conditions,
\begin{eqnarray}
{\bf M}^2_{R11}&=&\mp\left(\frac{s_{13}}{c_{13}}-\frac{c_{13}}{s_{13}}
\right)
{\bf M}^2_{R13}+{\bf M}^2_{R33}
\nonumber\\
{\bf M}^2_{R12}&=&\mp\frac{s_{13}}{c_{13}}{\bf M}^2_{R23}
\\
{\bf M}^2_{R22}&=&\mp\frac{s_{13}}{c_{13}}{\bf M}^2_{R13}
\mp\left(\frac{s_{23}}{c_{23}}-\frac{c_{23}}{s_{23}}\right)
\frac{1}{c_{13}}{\bf M}^2_{R23}
+{\bf M}^2_{R33}
\nonumber
\end{eqnarray}
which allow us to eliminate three parameters of the potential in favour
of the three angles in eq.~(\ref{Oangles}). These three parameters are
chosen as,
\begin{eqnarray}
\beta_1 &=& \frac{1}{2v_1^2}\left({\bf M}^2_{R11}-
\half\kappa v_2^2\frac{v_3}{v_1}\right)
\nonumber\\
\beta_2 &=& \frac{1}{v_1v_2}\left({\bf M}^2_{R12}+
\kappa v_2v_3\right)
\label{b1b2l1}\\
\lambda_1 &=& \frac{1}{2v_2^2}{\bf M}^2_{R22}
\nonumber
\end{eqnarray}
with the above expressions found  from eq.~(\ref{MR}). 

We do a scan of parameter space, imposing fermiophobia in the way just
described. The free parameters are varied according to,
\begin{eqnarray}
0.5\,{\mathrm{GeV}} <v_1< 1\,{\mathrm{TeV}}\,,
\qquad & v_2=246\,{\mathrm{GeV}}\,, &
\qquad v_3<0.35\,{\mathrm{GeV}}\,,\nonumber\\
0<\kappa<0.1\,,\qquad & |\beta_3|<4\,, & \qquad |\lambda_{2...5}|<4\,,
\label{fermphScan}\\
\theta_{12}=0\,,\qquad & 0<\theta_{13}<2\pi\,, & \qquad 
0<\theta_{23}<2\pi\,.
\nonumber
\end{eqnarray}
checking  that  $\beta_1$,  $\beta_2$,  and $\lambda_1$  have  all  an
absolute value smaller than 4.

%
%------------------------ FIGURE -------------------------------------
\begin{figure}[h!]
\centerline{
\protect
\vbox{
\epsfig{file=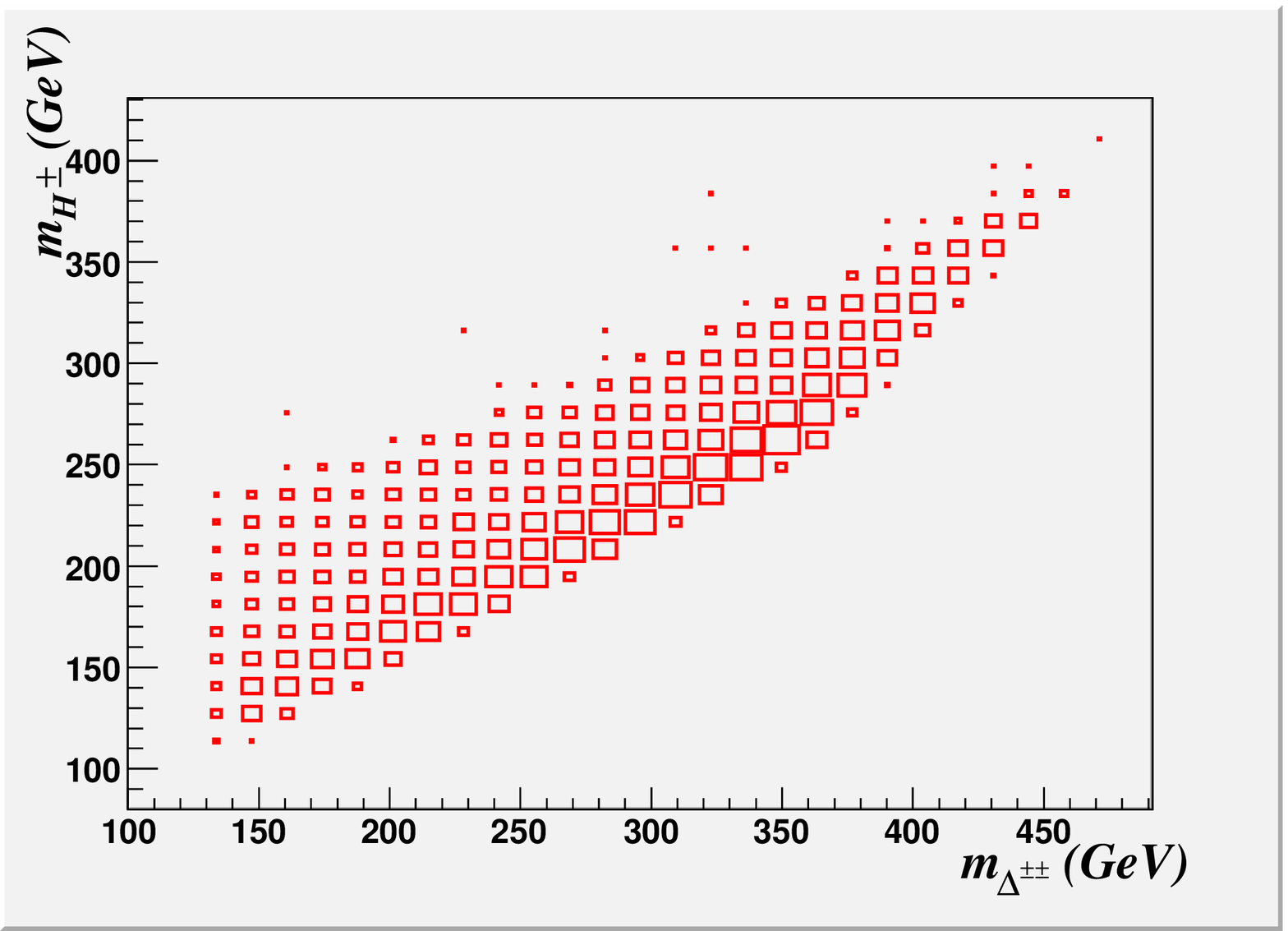,width=0.47\textwidth}
\epsfig{file=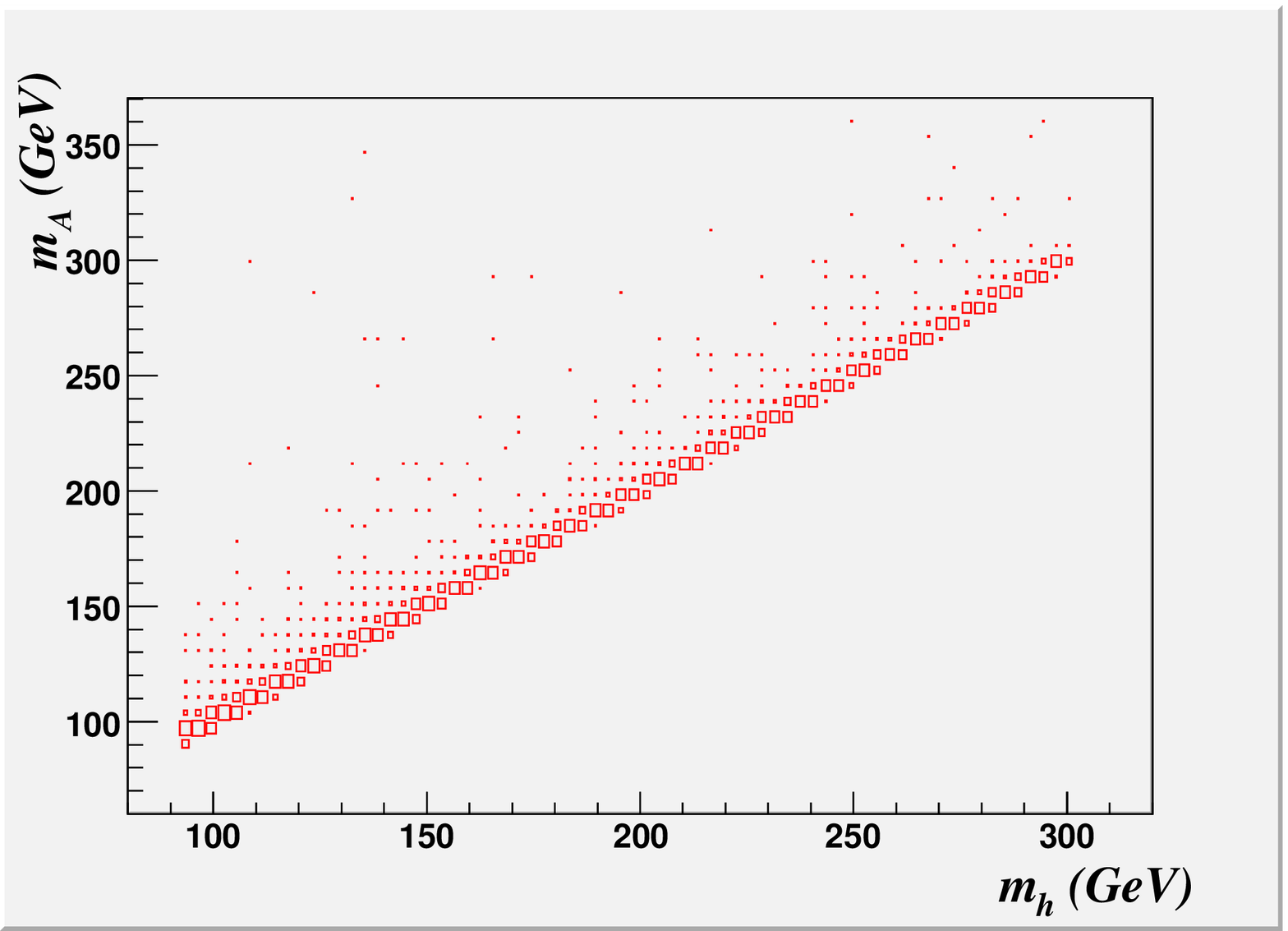,width=0.47\textwidth}
}}
\caption{\it Charged Higgs and  doubly-charged Higgs boson masses as a
  function of the fermiophobic Higgs mass.  }
\label{FERMIO_masses}
\end{figure}
%---------------------------------------------------------------------
%

In order to  study the masses of the scalars  in the pure fermiophobic
limit, in Fig.~\ref{FERMIO_masses} we show a scatter plot of $m_{H^+}$
and $m_{\Delta^{++}}$  (left frame), and  a scatter plot of  $m_A$ and
$m_{H_1}$  (right frame),  resulting from  the scan.   The correlation
between the singly and doubly charged Higgs boson masses is understood
from  the small  value  of the  triplet  vev, which  implies that  the
charged Higgs bosons satisfy,
\begin{eqnarray}
m_{H^+}^2 &\approx& m_A^2-\fourth\lambda_5v_2^2 \nonumber\\
m_{\Delta^{++}}^2 &\approx& m_A^2-\half\lambda_5v_2^2 \label{eq:mchargapp}
\end{eqnarray}
This  indicates that the  singly charged  Higgs boson  mass lies  in a
narrower region  than the doubly  charged Higgs mass, and  this effect
can be seen in the figures. Note that this implies,
\begin{equation}
m_{H^+}^2\approx\half(m_A^2+m_{\Delta^{++}}^2)
\label{mHmAmD}
\end{equation}
which is a very good approximation up to order ${\cal O}(v_3)$. 

The correlation between  the CP-odd Higgs mass $m_A$  and the lightest
CP-even   Higgs  mass   $m_{H_1}$   seen  in   the   right  frame   of
Fig.~\ref{FERMIO_masses} is explained  by inspecting the CP-even Higgs
boson  mass   matrix  in  eq.~(\ref{MR}),  and  the   CP-odd  mass  in
eq.~(\ref{ma}). Due to the hierarchy of vevs we see that,
\begin{equation}
({\bf M}_R^2)_{33} \approx m_A^2 \approx \frac{\kappa v_1v_2^2}{2v_3}
\sim m_{H_1}^2
\end{equation}
where the last relation comes from  the fact that most of the time the
CP-even  Higgs  mass  matrix  is  nearly  diagonal,  with  the  $({\bf
  M}_R^2)_{33}$ element being the smallest.

Another point  we wish to emphasize  here is the  relation between the
couplings of  the fermiophobic Higgs  boson to charged scalars  and to
gauge bosons.
%
%------------------------ FIGURE -------------------------------------
\begin{figure}[h!]
\centerline{\protect\vbox{\epsfig{file=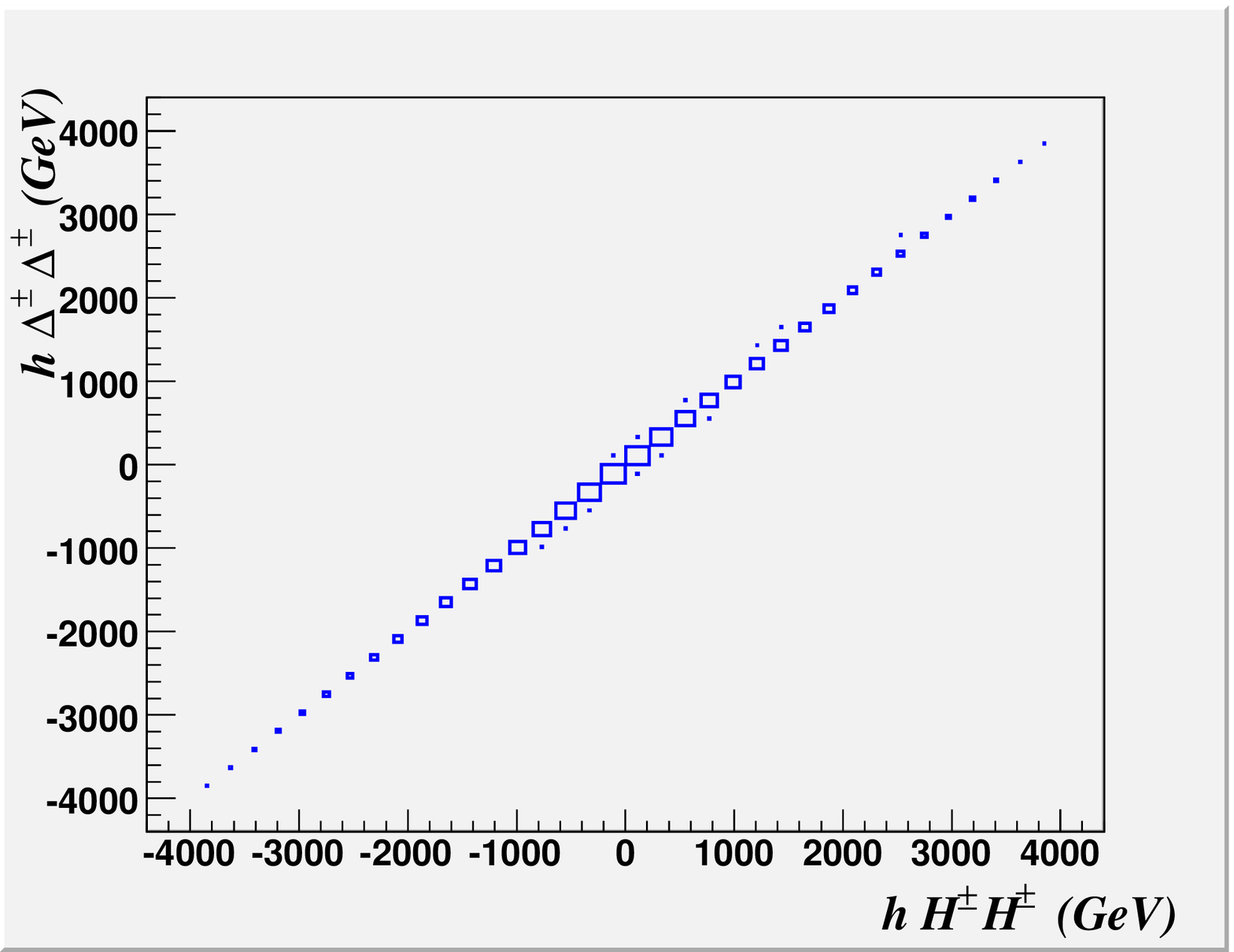,width=0.47\textwidth}
\epsfig{file=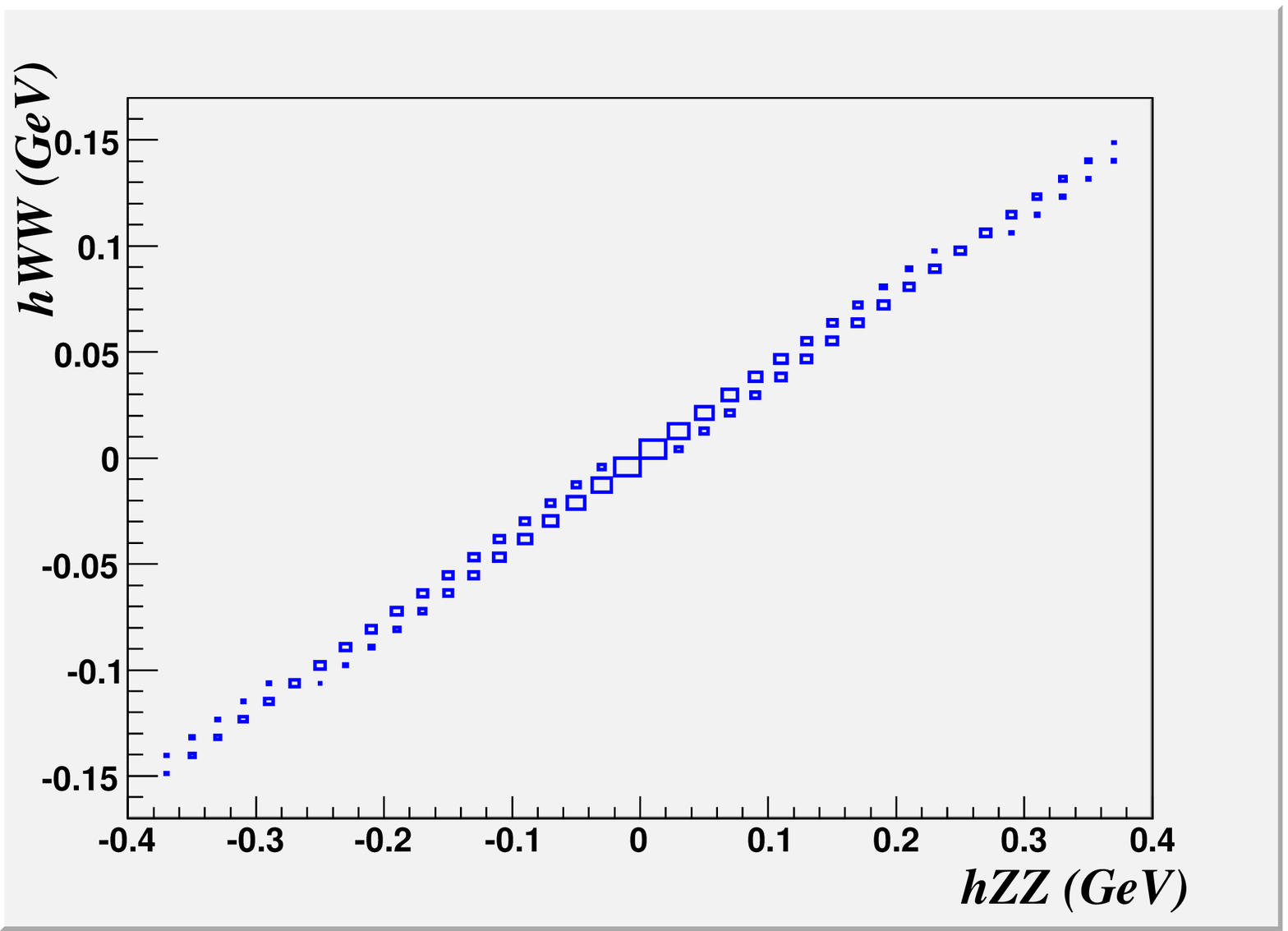,width=0.50\textwidth}}}
\caption{\it Fermiophobic  Higgs boson couplings to a  pair of charged
  Higgs (left) and to a pair of gauge bosons(right).  }
\label{CouplingsCargados_box}
\end{figure}
%---------------------------------------------------------------------
%
In  the left  frame  of Fig.~\ref{CouplingsCargados_box}  we show  the
relation  between the  couplings $h_f\Delta^{++}\Delta^{--}$  and $h_f
H^+H^-$  (see appendix~\ref{app:A}  for Feynman rules),  when we
vary  all  parameters  as  indicated  in  eq.~(\ref{fermphScan}).  The
couplings are clearly  proportional to each other, as  can be inferred
from the Feynman rules, which in the fermiophobic case satisfy,
\begin{equation}
\frac{\lambda(h_f\Delta^{++}\Delta^{--})}{\lambda(h_f H^+H^-)}
=1+{\cal{O}}(v_3)
\end{equation}
These couplings can have a magnitude as large as  4 TeV, although
in most of  the parameter space they are  $\lsim 300$ GeV.  Similarly,
in  the right  frame of  Fig.~\ref{CouplingsCargados_box} we  plot the
relation between the couplings $h_f W^+W^-$ and $h_f ZZ$.
\begin{equation}
\frac{\lambda(h_1     W^+    W^-)}{\lambda(h_1     Z     Z)}=    c_W^2
\frac{2v_3O^{13}_R+v_2O^{12}_R}{4v_3O^{13}_R+v_2O^{12}_R}
\label{RatioCoup}
\end{equation}
and if we take the exact fermiophobic limit we get,
\begin{equation}
\frac{\lambda(h_f W^+ W^-)}{\lambda(h_f Z Z)}
\longrightarrow\half c_W^2
\end{equation}
which is half  the value for a SM Higgs  boson.  This has implications
that will  be evaluated  in the next  sections.  Note that  this limit
changes   drastically   with   a   small  but   non-zero   value   for
$O^{12}_R$. Notice also  that the couplings $h_f W^+W^-$  and $h_f ZZ$
are much  smaller in  this fermiophobic limit  than the  equivalent SM
couplings.

\begin{figure}[h!]
\centerline{
\protect
\vbox{
\epsfig{file=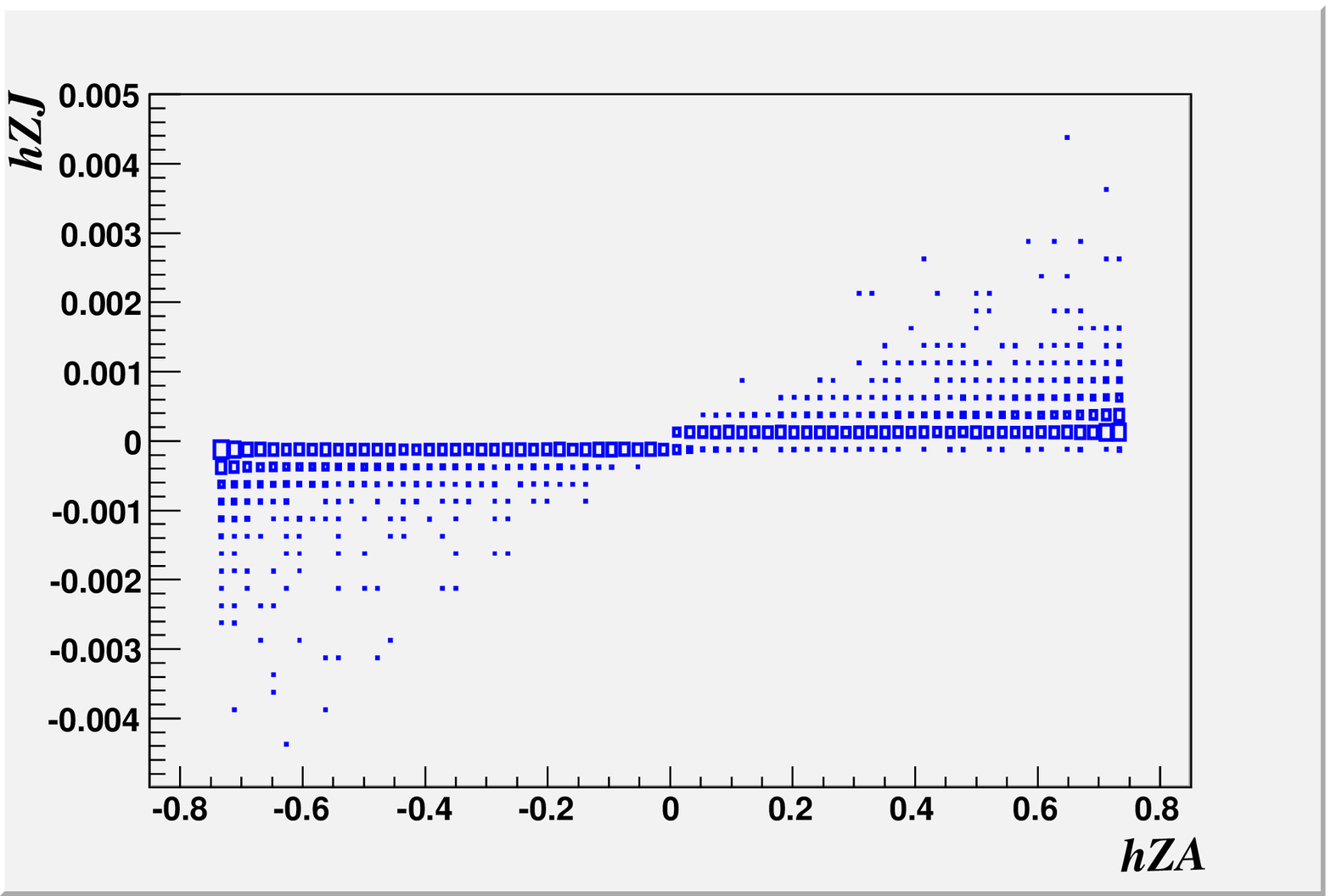,width=0.525\textwidth}
\epsfig{file=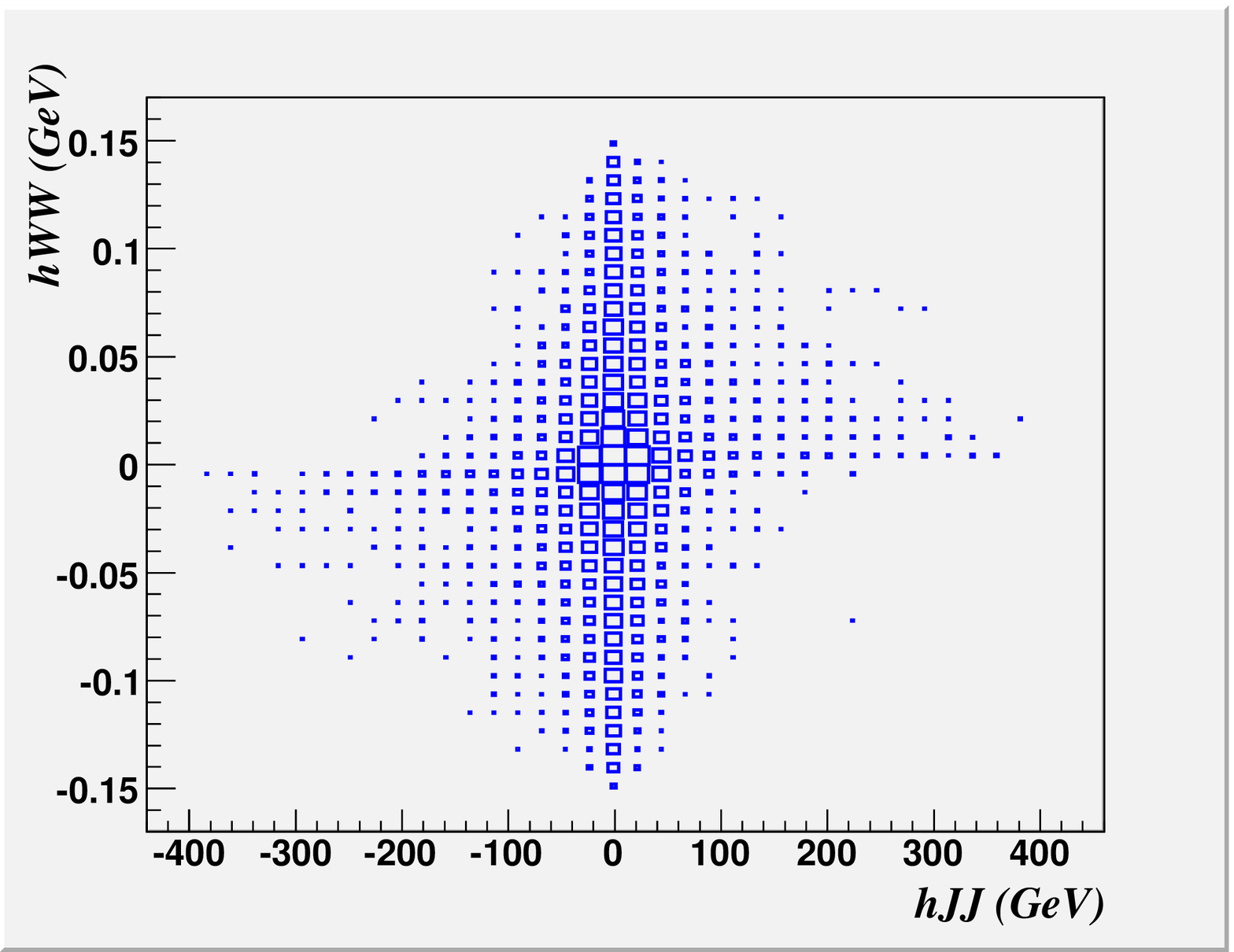,width=0.46\textwidth}}}
\caption{\it Relation  between couplings  with and without  Majoron in
  the Fermiophobic scenario.  }
\label{Couplings-Majoron}
\end{figure}

Within the same  scan of parameter space in  the fermiophobic scenario
given      by      eq.~(\ref{fermphScan})      we      explore      in
Fig.~\ref{Couplings-Majoron}  the  magnitude  of  couplings  with  and
without a  Majoron. In the left  frame of Fig.~\ref{Couplings-Majoron}
we have  the relation between  the $\lambda(h_f ZJ)$  and $\lambda(h_f
ZA)$ couplings. The $\lambda(h_f ZA)$ coupling can be large, but it is
not relevant for the decay of a fermiophobic Higgs, since $m_{h_f}$ is
rarely     larger     than      $m_Z+m_A$,     as     indicated     in
Fig.~\ref{FERMIO_masses}. On  the contrary, the  coupling $\lambda(h_f
ZJ)$ is  very small  and irrelevant for  production of  a fermiophobic
Higgs boson.  Nevertheless, the  decay $h_f\rightarrow ZJ$ is possible
and it is  characterized by missing energy. We  evaluate its branching
ratio in the next section.

%
%------------------------ FIGURE -------------------------------------
%---------------------------------------------------------------------
%

In the right frame of Fig.~\ref{Couplings-Majoron} we see the relation
between  the $\lambda(h_f  JJ)$ and  $\lambda(h_f WW)$  couplings. The
coupling $\lambda(h_f WW)$ is diminished  compared to the value in the
SM,   implying   that   the   decay  rate   $h_f\rightarrow   WW$   is
small. Nevertheless its branching  ratio, together with the comparable
BR for  $h_f\rightarrow ZZ$, will dominate unless  the invisible decay
$h_f\rightarrow JJ$ is large. This decay is controlled by the coupling
$\lambda(h_f   JJ)$,   also   displayed   in  the   right   frame   of
Fig.~\ref{Couplings-Majoron}. We see that  this coupling can be large,
in  which case  the  fermiophobic  Higgs will  be  invisible. In  this
scenario, one has to look for the second lightest Higgs boson.
%
%%%%%%%%%%%%%%%%%%%%%%%%%%%%%%%%%%%%%%%%%%%%%%%%%%%%%%%%%%%%%%%%%%%%%%
\subsection{Imposing Fermiophobia and Majoron Suppression}
%%%%%%%%%%%%%%%%%%%%%%%%%%%%%%%%%%%%%%%%%%%%%%%%%%%%%%%%%%%%%%%%%%%%%%
%
We  impose exact fermiophobia  and Majoron  suppression by  fixing the
value of  $\cos\theta_{13}=c_{13}=0$ in  eq.(\ref{Oangles}). The
diagonalizing matrix in this case is,
\begin{eqnarray}
O_R = \left[\begin{matrix}
0 & 0 & \pm 1 \cr 
-c_{23}s_{12} \mp s_{23}c_{12} & c_{23}c_{12} \mp s_{23}s_{12} & 0 \cr
s_{23}s_{12} \mp c_{23}c_{12}  & -s_{23}c_{12} \mp c_{23}s_{12} & 0
\end{matrix}\right]
=
\left[\begin{matrix}
0 & 0 & \pm 1 \cr 
-\sin\theta & \cos\theta & 0 \cr
\mp \cos\theta  & \mp \sin\theta & 0
\end{matrix}\right]
\end{eqnarray}
where $\pm$ corresponds to $\mathrm{sign}(s_{13})$, and $\theta \equiv
\theta_{12} \pm  \theta_{23}$ is introduced as  a new independent
  parameter.   The  last definition  indicates  that  only one  angle
controls the rotation matrix in this scenario.

The diagonalized  CP-even Higgs mass matrix  $({\bf M}^2_{R})_{diag} =
O_R {\bf M}^2_{R} O_R^T$, implies,
\begin{eqnarray}
{\bf M}^2_{R13}&=& 0
\nonumber\\
{\bf M}^2_{R23}&=& 0
\\
{\bf M}^2_{R12}
&=&
\frac{1}{2} \big( {\bf M}^2_{R11}-{\bf M}^2_{R22} \big) \tan 2\theta 
\nonumber
\end{eqnarray}
From these  equations we eliminate  the following parameters  from the
list of independent parameters given in eq.~(\ref{scanDef}),
\begin{eqnarray}
\beta_3 &=& \frac{\kappa v_2^2}{2v_1v_3}
\cr
\kappa &=& (\lambda_3+\lambda_5) \frac{v_3}{v_1}
\cr
\beta_2 &=& \kappa\frac{v_3}{v_1} + \frac{1}{2v_1v_2}
\Big( 2\beta_1 v_1^2+\frac{1}{2}\kappa v_2^2\frac{v_3}{v_1}-2
\lambda_1 v_2^2 \Big)
\tan 2\theta
\label{ConstraintsFMs}
\end{eqnarray}
We do a scan of the parameter space, imposing fermiophobia and Majoron
suppression.  The free parameters are varied according to,

\begin{eqnarray}
0.5\,{\mathrm{GeV}} <v_1< 1\,{\mathrm{TeV}}\,,
\quad & v_2=246\,{\mathrm{GeV}}\,, &
\quad v_3<0.35\,{\mathrm{GeV}}\,,\nonumber\\
|\beta_1|<4\,, & \qquad |\lambda_{1...5}|<4\,,\qquad & 0<\theta<2\pi\,.
\label{fermphMSupScan}
\end{eqnarray}
checking that  $\beta_2$, $\beta_3$  and $\kappa$ have  values smaller
than 4.

%
%------------------------ FIGURE -------------------------------------
\begin{figure}[h!]
\centerline{
\protect
\vbox{
\epsfig{file=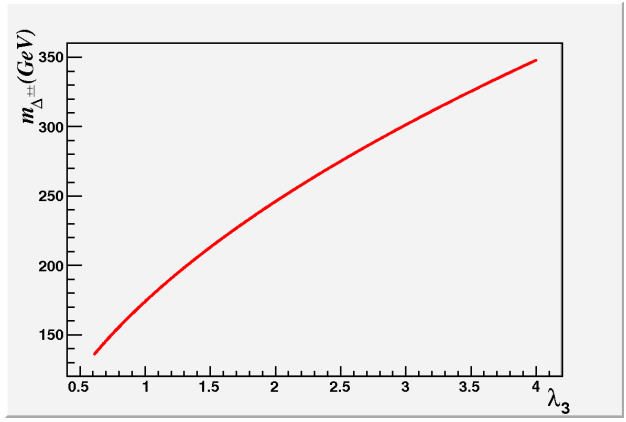,width=0.5\textwidth}}}
\caption{\it  Doubly  charged  Higgs  mass dependence  on  $\lambda_3$
  coupling in the Fermiophobic and Majoron suppressed scenario.  }
\label{MassDeltaL3}
\end{figure}
%---------------------------------------------------------------------
%

First we notice the clear  dependence of the doubly charged Higgs mass
$m_{\Delta^{++}}$        on        $\lambda_3$,        shown        in
Fig.~\ref{MassDeltaL3}. This dependence  is easily understood from the
expression  for the  mass $m_{\Delta^{++}}$  in eq.~(\ref{DeltaMass}),
which   after  replacing   $\kappa$   from  eq.~(\ref{ConstraintsFMs})
transforms into,
\begin{equation}
m_{\Delta^{++}}^2 \approx \frac{1}{2}\lambda_3 v_2^2 -\lambda_4 v_3^2
\end{equation}
with  the  equality holding  in  the  exact  fermiophobic and  Majoron
suppression scenario. Since $v_3\ll v_2$, this relation would enable a
direct determination  of $\lambda_3$ from experiments.  Note also that
the relation  in eq.~(\ref{mHmAmD}) is valid also  in the fermiophobic
plus Majoron suppression  scenario (since it is a  special case of the
fermiophobic case),  thus providing a measurement  that could validate
or refute the model.

%
%------------------------ FIGURE -------------------------------------
\begin{figure}[h!]
\centerline{
\protect
\vbox{
\epsfig{file=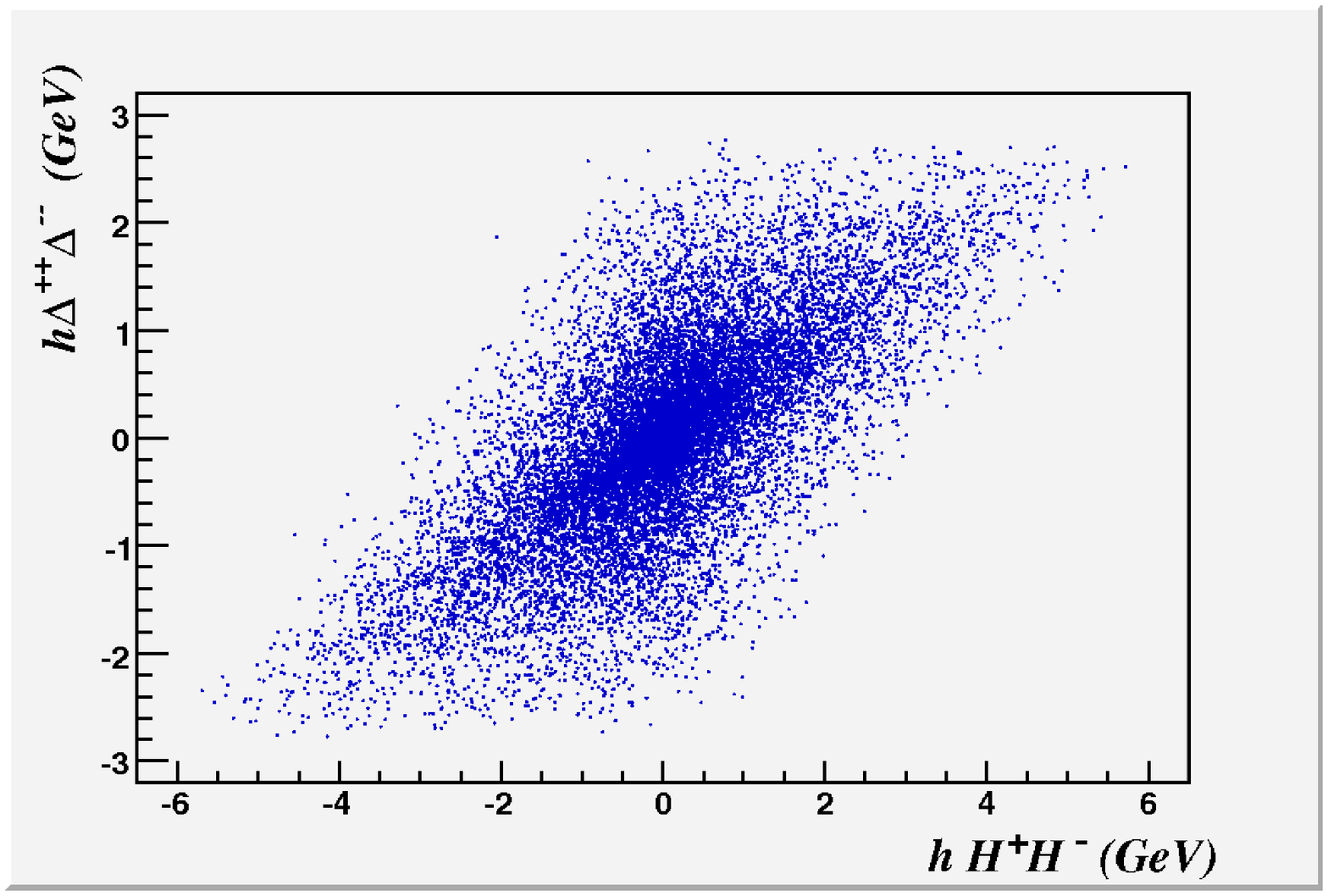,width=0.49\textwidth}
\epsfig{file=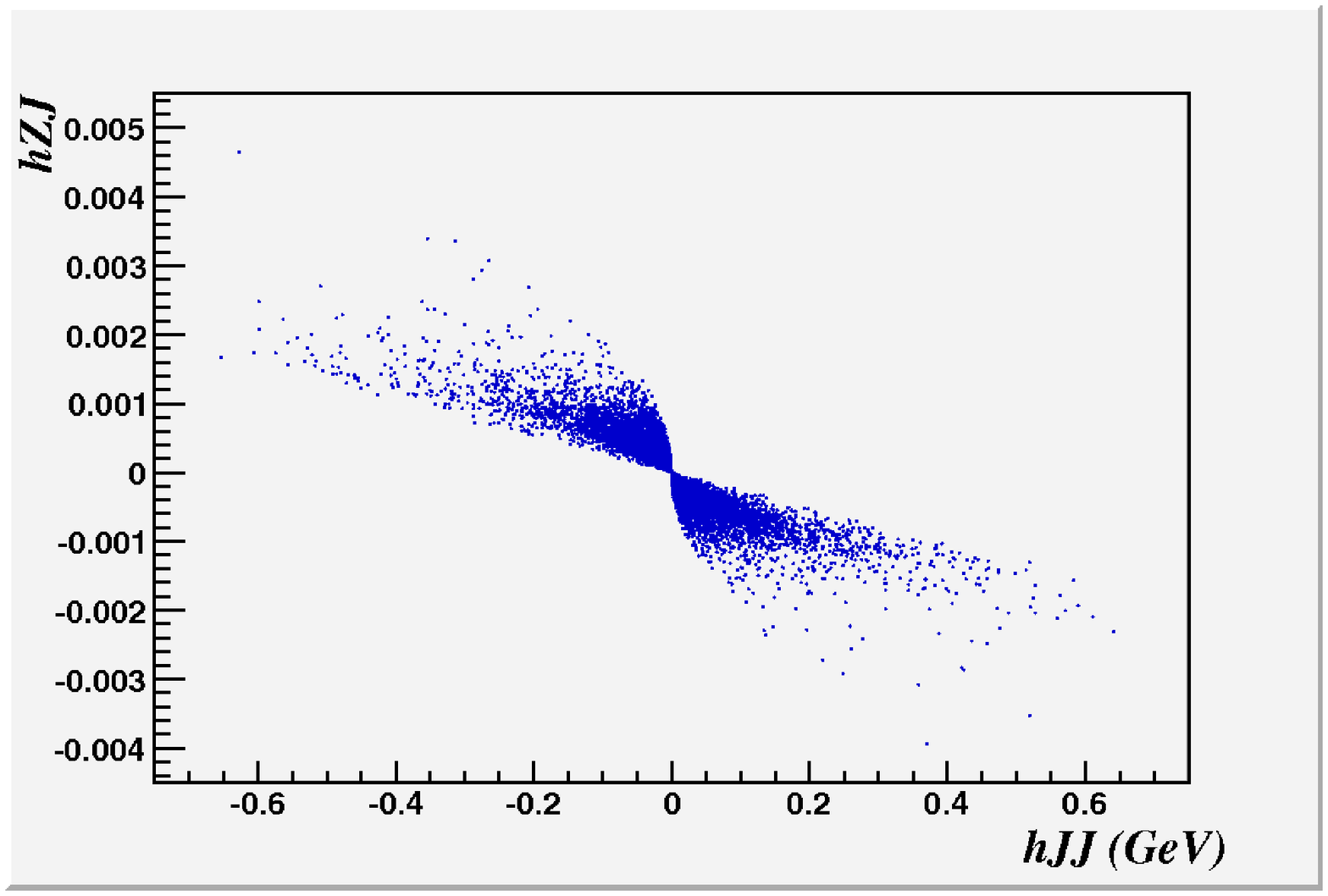,width=0.49\textwidth}
}}
\caption{\it Relation  between couplings  with and without  Majoron in
  the Fermiophobic and Majoron suppressed scenario.  }
\label{Couplings-Majoron3}
\end{figure}
%---------------------------------------------------------------------
%

In  the  left  frame  of  Fig.~\ref{Couplings-Majoron3}  we  plot  the
relation between the fermiophobic Higgs couplings to a pair of charged
Higgs  $\lambda(h_fH^+H^-)$ and  to  a pair  of  doubly charged  Higgs
bosons $\lambda(h_f\Delta^{++}\Delta^{--})$.  Clearly, the suppression
of   the  Majoron   component  in   the  fermiophobic   Higgs  reduces
dramatically the value of these couplings (compare with the left frame
of  Fig.~\ref{CouplingsCargados_box}),  which  are important  for  the
decay rate for the $h_f\rightarrow\gamma\gamma$ mode.

In the right frame of Fig.~\ref{Couplings-Majoron3} we study couplings
that involve  one or  two Majorons.  To be more  specific, we  see the
relation between  $\lambda(h_f ZJ)$  and $\lambda(h_f JJ)$,  the first
one corresponding to  $h_f$ decay with missing energy,  and the second
one  corresponding to  an invisible  decay. The  coupling $\lambda(h_f
ZJ)$ maintains its magnitude  compared with the previous scenario, but
$\lambda(h_f  JJ)$ is  much  smaller, an  expected  effect since  this
scenario is defined by a null singlet component in $h_f$.
%
%%%%%%%%%%%%%%%%%%%%%%%%%%%%%%%%%%%%%%%%%%%%%%%%%%%%%%%%%%%%%%%%%%%%%%
\section{Fermiophobic Higgs Boson Decays}
\label{sec:fdecay}
%%%%%%%%%%%%%%%%%%%%%%%%%%%%%%%%%%%%%%%%%%%%%%%%%%%%%%%%%%%%%%%%%%%%%%
%
A fermiophobic  Higgs boson  will have four  main decay modes,  two of
them into a pair of  massive gauge bosons, $h_f\rightarrow W^+W^-$ and
$h_f\rightarrow  ZZ$, where  one or  two of  the gauge  bosons  may be
off-shell depending on  the Higgs mass, and the other  two into one or
two photons $h_f\rightarrow\gamma\gamma$ and $h_f\rightarrow\gamma Z$,
decays which are generated at one-loop \cite{Phalen:2006ga}.

In  the  later  case,  $W$  gauge boson  contribute  to  the  one-loop
generated decay with the graphs,
%
%------------------------ FIGURE -------------------------------------
\begin{center}
\vspace{-85pt} \hfill \\
\begin{picture}(300,140)(0,23) % y_2 controls equation position
\DashLine(20,30)(60,30){3}
\Photon(60,30)(90,60){3}{6.5}
\Photon(90,0)(60,30){3}{6.5}
\Photon(90,60)(90,0){3}{8}
\Photon(90,60)(130,60){3}{6.5}
\Photon(90,0)(130,0){3}{6.5}
\Text(20,40)[]{$h_f$}
\Text(140,60)[]{$\gamma$}
\Text(144,0)[]{$\gamma, Z$}
\Text(107,30)[]{$W^{\pm}$}
\DashLine(170,30)(210,30){3}
\PhotonArc(230,30)(20,0,360){3}{18}
\Photon(250,30)(280,60){3}{6.5}
\Photon(250,30)(280,0){3}{6.5}
\Text(170,40)[]{$h_f$}
\Text(290,60)[]{$\gamma$}
\Text(294,0)[]{$\gamma, Z$}
\Text(230,60)[]{$W^{\pm}$}
\end{picture}  %after this line goes the equation
\vspace{30pt} \hfill \\
\end{center}
%---------------------------------------------------------------------
%
present already in the SM.  These graphs are complemented in our model
with singly and doubly charged Higgs bosons,
%
%------------------------ FIGURE -------------------------------------
\begin{center}
\vspace{-10pt} \hfill \\
\begin{picture}(300,140)(0,23) % y_2 controls equation position
\DashLine(20,120)(60,120){3}
\DashLine(60,120)(90,150){3}
\DashLine(90,90)(60,120){3}
\DashLine(90,150)(90,90){3}
\Photon(90,150)(130,150){3}{6.5}
\Photon(90,90)(130,90){3}{6.5}
\Text(20,130)[]{$h_f$}
\Text(140,150)[]{$\gamma$}
\Text(144,90)[]{$\gamma, Z$}
\Text(115,122)[]{$H^+, \Delta^{++}$}
\DashLine(170,120)(210,120){3}
\DashCArc(230,120)(20,0,360){3}
\Photon(250,120)(280,150){3}{6.5}
\Photon(250,120)(280,90){3}{6.5}
\Text(170,130)[]{$h_f$}
\Text(290,150)[]{$\gamma$}
\Text(294,90)[]{$\gamma, Z$}
\Text(233,150)[]{$H^+, \Delta^{++}$}
\end{picture}  %after this line goes the equation
\vspace{-40pt} \hfill \\
\end{center}
%---------------------------------------------------------------------
%
The decay rate for $h_f\to \gamma\gamma$ is given by,
\begin{equation}
\Gamma(h_f\to \gamma\gamma)={\alpha^2g^2\over 1024\pi^3}
{m^3_{h_f}\over m_W^2} 
|F_0(\tau_H) \, \tilde g_{hHH}+
4 F_0(\tau_\Delta) \, \tilde g_{h\Delta\Delta} -
F_1(\tau_W) \, \tilde g_{hWW}|^2 
\label{withscalarloops}
\end{equation}
where  $F_0$ and  $F_1$ are  loop functions  associated to  scalar and
vector   bosons,   and   explicit   expressions  can   be   found   in
ref.~\cite{Gunion:1989we}.      They     are     a     function     of
$\tau_i=4m_i^2/m_f^2$,  where  $i=H,  \Delta, W$.   The  dimensionless
couplings $\tilde g_{hHH}$ and $\tilde g_{h\Delta\Delta}$ are,
\begin{equation}
\tilde g_{hHH}= \frac{m_W}{gm_{H^+}^2}\lambda(h_f H^+H^-)
\,,\qquad
\tilde g_{h\Delta\Delta}= 
\frac{m_W}{gm_{\Delta^{++}}^2}\lambda(h_f \Delta^{++}\Delta^{--})
\label{trilinear_coup1}
\end{equation}
where in the fermiophobic case with $O^{11}_R\ne0$ we find,
\begin{equation}
\lambda(h_f H^+H^-)\approx\lambda(h_f \Delta^{++}\Delta^{--})
\approx O^{11}_R\beta_3v_1
\end{equation}
up to terms of ${\cal{O}}(v_3)$. In the fermiophobic case with Majoron
suppression  scenario,   where  $O^{11}_R=0$,  the   terms  that  were
sub-leading in the previous case, dominate now,
\begin{eqnarray}
\lambda(h_f H^+H^-) &\approx& 2\lambda_2 v_3
\cr
\lambda(h_f \Delta^{++}\Delta^{--}) &\approx& 
2(\lambda_2+\lambda_4-\frac{1}{2}\lambda_5) v_3
\end{eqnarray}
which translate into a diminished  influence of the charged scalars in
the   decay   rate   for   $h_f\rightarrow\gamma\gamma$.   The   other
contribution  to $h_f\rightarrow\gamma\gamma$  is from  the  $W$ loop,
controlled  by  the  dimensionless  coupling $\tilde  g_{hWW}$,  which
satisfies in both scenarios:
\begin{equation}
\tilde g_{hWW}=\frac{1}{g\,m_W}\lambda(h_f W^+W^-)=
\frac{g O^{13}_R v_3}{m_W}
\end{equation}
This implies that in the case of fermiophobia with Majoron suppression
all      the      contributions       from      $W$      loops      to
$\Gamma(h_f\rightarrow\gamma\gamma)$ are suppressed.
%
%%%%%%%%%%%%%%%%%%%%%%%%%%%%%%%%%%%%%%%%%%%%%%%%%%%%%%%%%%%%%%%%%%%%%%
\section{Phenomenology of a HTM Fermiophobic Higgs Boson}
\label{sec:pheno}
%%%%%%%%%%%%%%%%%%%%%%%%%%%%%%%%%%%%%%%%%%%%%%%%%%%%%%%%%%%%%%%%%%%%%%
%
In the SM the main Higgs boson decay channels are $h \to b\bar{b}$ and
$h\rightarrow \tau^+\tau^-$  (for $m_h\le 130$  GeV).  In fermiophobic
models though, like the ones based on  the 2HDM Type I or the HTM, the
fermionic decays are suppressed. As a consequence, decay channels into
gauge bosons  become the most  important ones, including  the one-loop
generated decays  $h\to \gamma \gamma$ and  $h \to \gamma  Z$. Here we
study the decay rates of  a fermiophobic Higgs boson into gauge bosons
in our  HTM, including decay modes  with Majorons. For  an analysis in
the 2HDM see \cite{Akeroyd:2007yh,Arhrib:2008pw}.
%
%------------------------ FIGURE -------------------------------------
\begin{figure}[!h]
\begin{center}
\begin{tabular}{cc}
\includegraphics[angle=0,width=8.3cm]{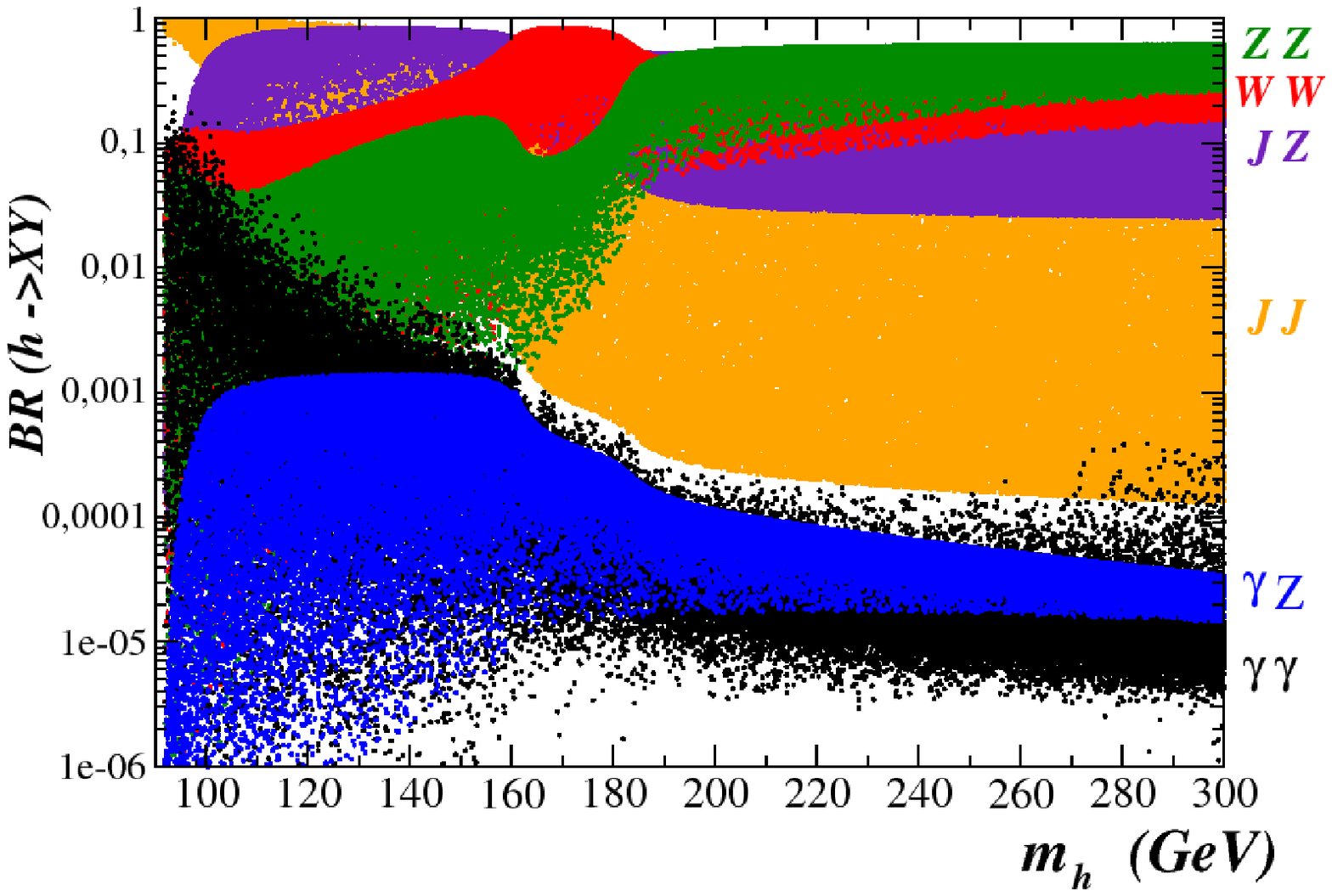}
&
\includegraphics[angle=0,width=8.1cm]{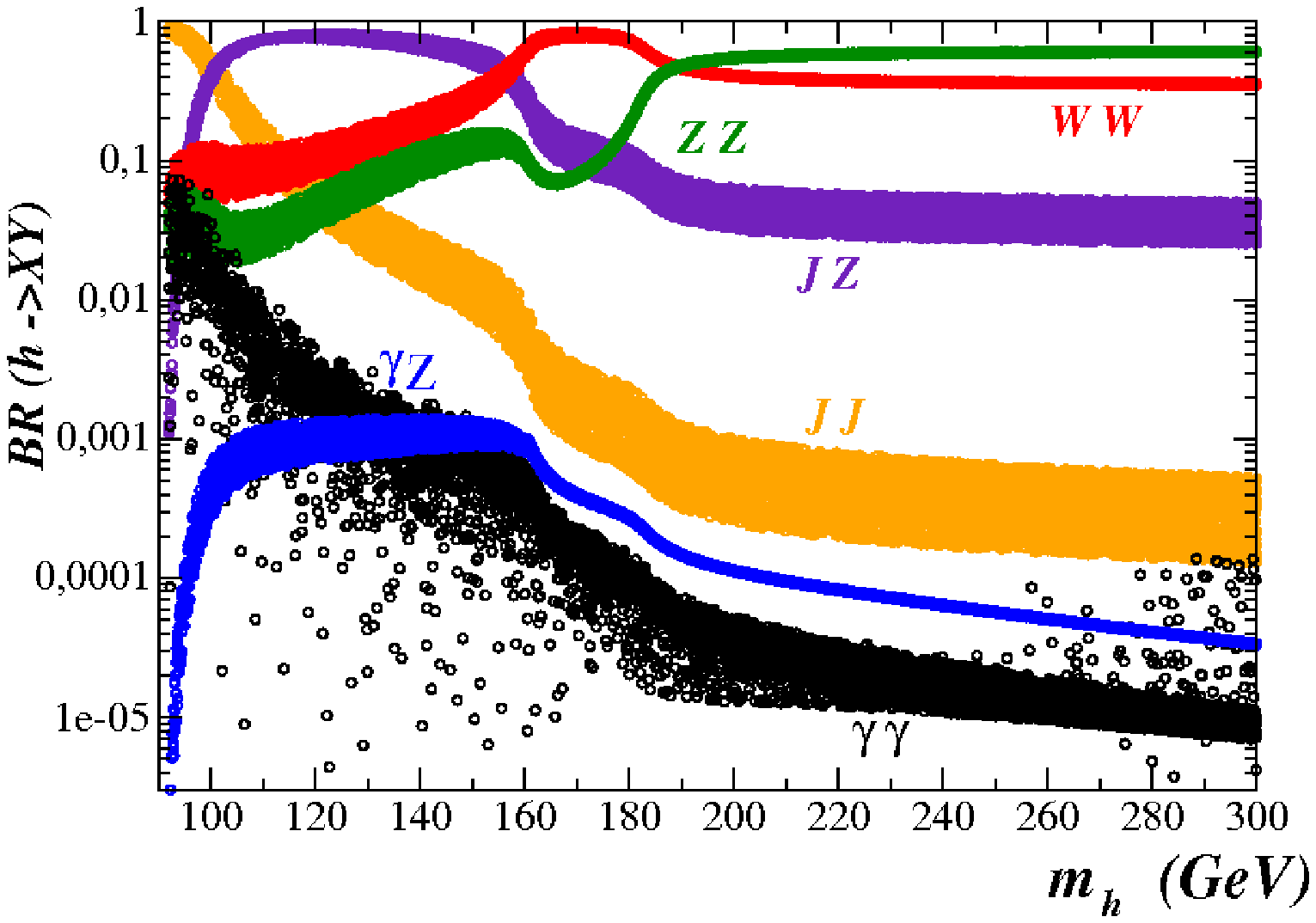}
\end{tabular}
\end{center}
\caption{\it  Branching  ratios for  the  two-body fermiophobic  Higgs
  decays  $h_f \to  XY$ as  a function  of $m_f$  in  the fermiophobic
  scenario  with the  Majoron suppression  approximation. In  the left
  frame parameters  are varied freely,  while in the right  frame they
  are restricted as indicated in the text.
\label{BRmh}}
\end{figure}
%---------------------------------------------------------------------
%
In the  left frame of  Fig.~\ref{BRmh} we show the  fermiophobic Higgs
boson  branching ratios  $B(h_f \to  XY)$ as  a function  of  its mass
$m_f$, which  include four  gauge bosons modes:  $\gamma\gamma$, $ZZ$,
$WW$, and  $\gamma Z$, and  modes with one  or two Majorons:  $JJ$ and
$JZ$.  We randomly  vary the parameters in the  potential as indicated
in  eq.~(\ref{fermphScan}), imposing  exact fermiophobia  with Majoron
suppression via eq.~(\ref{ConstraintsFMs}).  We also use the following
experimental restrictions for the Higgs masses,
\[ m_{H^\pm} > 80 \, \textnormal{GeV,} \,\,\,\,\,\,\,\,
   m_{\Delta^{\pm \pm}} > 136 \, \textnormal{GeV,} \,\,\,\,\,\,\,\,
 m_h > 90 \, \textnormal{GeV} \,\,\,\,\,\,\,  \textnormal{and} 
\,\,\,\,\,\,\, m_A > 90 \, \textnormal{GeV}. \] 
We highlight from the  left frame of Fig.~\ref{BRmh} three distinctive
features: (i)  decay modes  with Majorons are  very important  for low
Higgs  masses; (ii)  decay modes  with  two massive  gauge bosons  can
dominate for large masses,  with a distinctive ratio $B(h_f\rightarrow
WW)/B(h_f\rightarrow  ZZ)$ as  compared with  the SM;  (iii) radiative
decays are suppressed with the exception of $\gamma\gamma$ at very low
masses.

In the  right frame of Fig.~\ref{BRmh}  we restrict a few  of the free
parameters  in  order to  better  visualize  the  above features.  The
dispersion of allowed points in parameter space is reduced by imposing
$700<v_1<1000$ GeV and $\beta_3<0.3$.

In the  case of  the aforementioned  feature (i), we  see in  the left
frame that the  decay mode $h_f\rightarrow JJ$ can  dominate at masses
$m_f<160$ GeV, while it  becomes smaller than $10\%$ for $m_f\gsim110$
GeV  in  the  restricted case  in  the  right  frame.  This  decay  is
invisible, since the Majoron  escapes detection. Its effect would show
up as a diminished production  cross section for the visible decays of
$h_f$. The decay mode $h_f\rightarrow  JZ$ is also very important, and
manifests  itself as  a $Z$  decaying into  two fermions  plus missing
energy. In the restricted case in the right frame, $h_f\rightarrow JZ$
dominates for masses $102<m_f<155$ GeV, being reduced to a few percent
for large Higgs masses.

Regarding feature  (ii), both decay modes $h_f\rightarrow  WW, ZZ$ are
above $5\%$  for masses  $m_f>190$ GeV, while  in the  restricted case
(right frame) the two branching ratios are larger than $5\%$
%\textbf{largely dominates}
for even smaller  masses ($m_f>130$ GeV). It is  also very interesting
to note that the decay  $h_f\rightarrow ZZ$ can have a branching ratio
larger  than  $h_f\rightarrow  WW$,  in  contrast with  the  SM  where
$B(h_f\rightarrow WW)/B(h_f\rightarrow ZZ)\approx 2.5$. This situation
appear for  $m_f>190$ GeV in  the right frame of  Fig.~\ref{BRmh}.  In
the restricted  case, the  two branching ratios  interchange dominance
around $m_f\sim190$ GeV. This is a very important feature of the model
since it can differentiate it  from other models.  The reason for this
behavior lies in the Higgs boson couplings to gauge bosons, and can be
understood from eq.~(\ref{RatioCoup}).

For feature (iii) we see in the left frame of Fig.~\ref{BRmh} that the
radiatively   generated  decay  $h_f\rightarrow\gamma\gamma$   can  be
potentially important at the LHC ($\gsim 6\%$) only for very low Higgs
masses $m_h\lsim 107$ GeV, while the $Z\gamma$ mode is irrelevant with
a  largest value  of  $\sim 0.1\%$  at  intermediate masses  $105\lsim
m_f\lsim 160$  GeV. These results can  also be seen  in the restricted
case in the right frame.

As we  mentioned, one  characteristic of our  model is that  the ratio
$B(h_f\rightarrow ZZ)/B(h_f\rightarrow WW)$  can be larger than unity,
while  in  fermiophobic  2HDM  and  in  the SM,  it  is  smaller  than
unity.  From eq.~(\ref{RatioCoup})  we  see that  if  the triplet  vev
vanishes  we recover  the  SM ratio.  On  the other  hand, with  exact
fermiophobia where $O_R^{12}=0$ we  get an inverted ratio with respect
to the SM  one. The obvious question is how  much deviation from exact
fermiophobia is needed to reestablish the SM limit. A similar question
is by  how much  the visible branching  ratios change when  we deviate
from  the scenario  of exact  Majoron suppression.   We  explore these
issues in the following figures.

%
%------------------------ TABLE  -------------------------------------
\begin{table}
\begin{center}
\caption{HTM parameters for four scenarios.}
\bigskip
\begin{minipage}[t]{0.8\textwidth}
\begin{tabular}{ccccccc}
\hline
Parameter & Scenario 1 & Scenario 2 & Scenario 3 & Scenario 4 & Scenario 5 & Units \\
\hline \hline
$v_1$ &639.54&931.26&919.58&950.94&862.93&GeV\\
$v_3$ &0.29&0.13&0.026&0.34&0.072&GeV\\
$\lambda_{1}$ &0.85 &2.53&2.81& 1.78&2.24&-\\
$\lambda_{2}$ &3.98 &3.45&2.86&0.72&-0.86&-\\
$\lambda_{3}$ & 0.68 &2.90&1.81&1.11&1.36&-\\
$\lambda_{4}$ & 2.72 &2.54&0.90&-1.68&-0.38&-\\
$\lambda_{5}$  & -0.37 &-2.62&-1.52&0.87&1.10&-\\
$\beta_{1}$  & 0.45 &0.17&0.20& 0.27&3.58&-\\
$\tan (2\theta)$& -0.3 & -1.25 & 1.1 & -0.34 & -0.015& -\\
\hline
$m_h$&96.55&92.40&95.18&244.78&272.33&GeV\\
$m_{H^{\pm}}$&122.18&219.57&178.89&216.06&239.98&GeV\\
$m_{\Delta^{\pm \pm}}$&143.29&296.46&234.40&182.89&202.53&GeV\\
$m_A$   &96.44&94.27&94.81&244.78&272.33&GeV\\
\hline
$hH^{\pm}H^{\mp}$ &-3.98&1.97&-2.33&0.95&-0.26&GeV\\
$h\Delta^{\pm\pm}\Delta^{\mp\mp}$ &-2.30&0.93&-0.14&-0.49&-0.12&GeV\\
$hWW$
&-0.12&$0.57\times10^{-1}$&$-0.11\times10^{-1}$&$-0.14$&$3.05\times10^{-2}
$&GeV\\
$hZZ$ &-0.31&0.15&$-0.27\times10^{-1}$&$-0.36$&$7.78\times10^{-2}$&GeV\\
$hJJ$
&$-0.33\times10^{-2}$&$0.66\times10^{-3}$&$-0.14\times10^{-3}$&$-1.12\times10^{
-2}$&$3.60\times10^{-3}$&GeV\\
%$hZA$ &0.73&-0.73&0.73&0.73&-0.73&-\\
$hZJ$
&$0.33\times10^{-3}$&-$0.11\times10^{-3}$&$0.21\times10^{-4}$&
$2.61\times10^{-4} $&$-6.15\times10^{-5} $&-\\
\hline \hline \label{tab:params}
\end{tabular}
\end{minipage}
\end{center}
\end{table}
%---------------------------------------------------------------------

We  define  five  scenarios  in  order to  explore  the  effects  from
deviating from exact fermiophobia  and Majoron suppression. The choice
of parameters is random with  the following exception: the first three
scenarios are chosen  such that the decay $h_f\rightarrow\gamma\gamma$
is enhanced, and  the last two are chosen  such that $B(h_f\rightarrow
ZZ)>B(h_f\rightarrow WW)$  for large Higgs masses, in  contrast to the
SM  prediction.   

The first  three scenarios are  characterized by a  small fermiophobic
Higgs mass,  while the  last two by  a large one.   Another difference
between  them is  that in  the  first three  scenarios $m_{H^{\pm}}  <
m_{\Delta^{\pm \pm}}$ as opposed to  the last two where $m_{H^{\pm}} >
m_{\Delta^{\pm   \pm}}$.    This  can   be   easily  explained   using
eq.(\ref{eq:mchargapp}).   The  five scenarios  are  defined in  Table
\ref{tab:params}.

Each scenario,  defined by  $O_R^{12}=O_R^{11}=0$, is analyzed  in the
next  five figures.  In  the left  frames  we explore  the effects  of
deviation  from exact  fermiophobia, taking  $O_R^{12}\gsim0$.  In the
right frames  we explore the  effects of deviation from  exact Majoron
suppression, taking $O_R^{11}\gsim0$.
%
%------------------------ FIGURE -------------------------------------
\begin{figure}[!ht]
\begin{center}
\begin{tabular}{cc}
\includegraphics[angle=0,width=8.1cm]{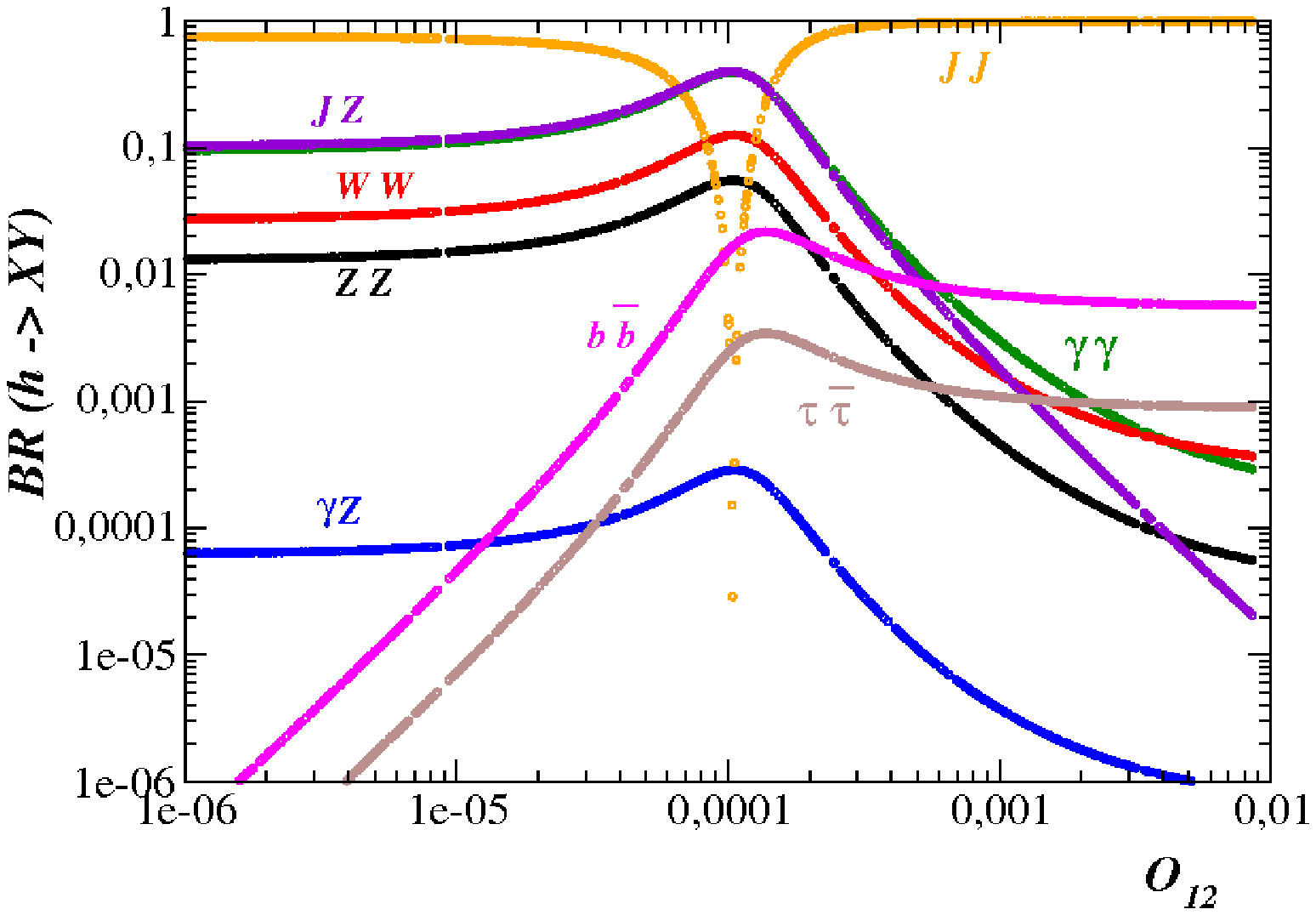}
&
\includegraphics[angle=0,width=8.3cm]{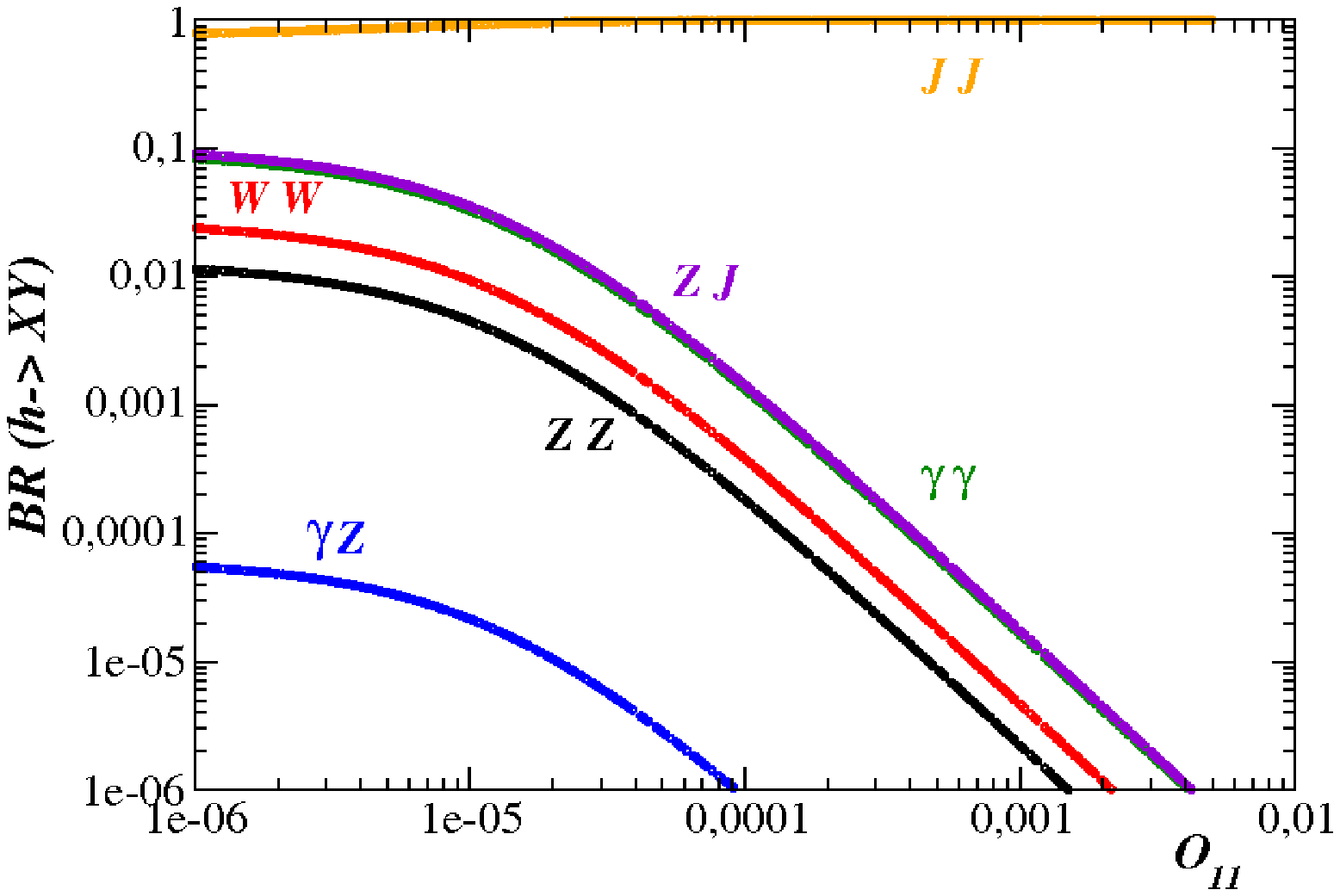}
\end{tabular}
\end{center}
\caption{\it Higgs  branching ratios for scenario  1, with $m_f=96.55$
  GeV.   Deviation from  exact  fermiophobia in  the  left frame,  and
  deviation from exact Majoron suppression in the right frame.
  \label{pheno:BR1}}
\end{figure}
%---------------------------------------------------------------------
%
In Fig.~\ref{pheno:BR1} we consider scenario 1. The enhancement of the
decay  $h_f\rightarrow\gamma\gamma$ is  achieved by  choosing  a light
Higgs boson, which  in scenario 1 is $m_h=96.55$  GeV. The relevant BR
are  $B(h_f\rightarrow\gamma\gamma)$ and  $B(h_f\rightarrow  JZ)$, and
they  are   close  to  $10\%$  for  exact   fermiophobia  and  Majoron
suppression,   with   the   invisible  mode   $B(h_f\rightarrow   JJ)$
dominating.   Deviation from exact  fermiophobia is  seen in  the left
frame: both $B(h_f\rightarrow\gamma\gamma)$ and $B(h_f\rightarrow JZ)$
grow  up  to  $40\%$  because  the $h_f  JJ$  coupling  vanishes  when
$O_R^{12}\approx  0.0001$.  After  that, they  decrease  sharply.  The
effect  of deviation  from exact  Majoron suppression  is seen  in the
right frame: all visible  decay modes diminish rapidly with increasing
$O_R^{11}$.
%
%------------------------ FIGURE -------------------------------------
\begin{figure}[!ht]
\begin{center}
\begin{tabular}{cc}
\includegraphics[angle=0,width=8.1cm]{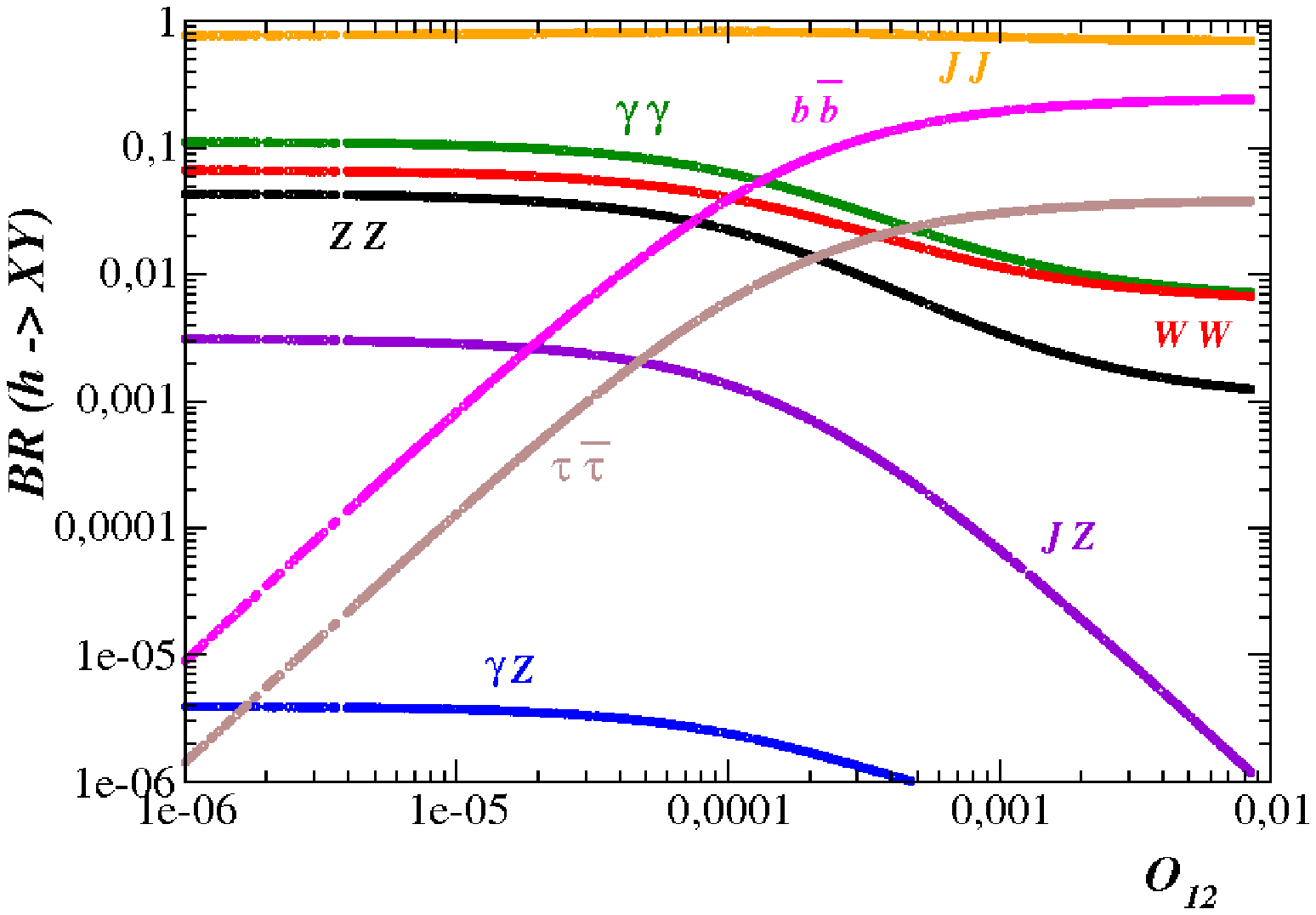}
&
\includegraphics[angle=0,width=8.3cm]{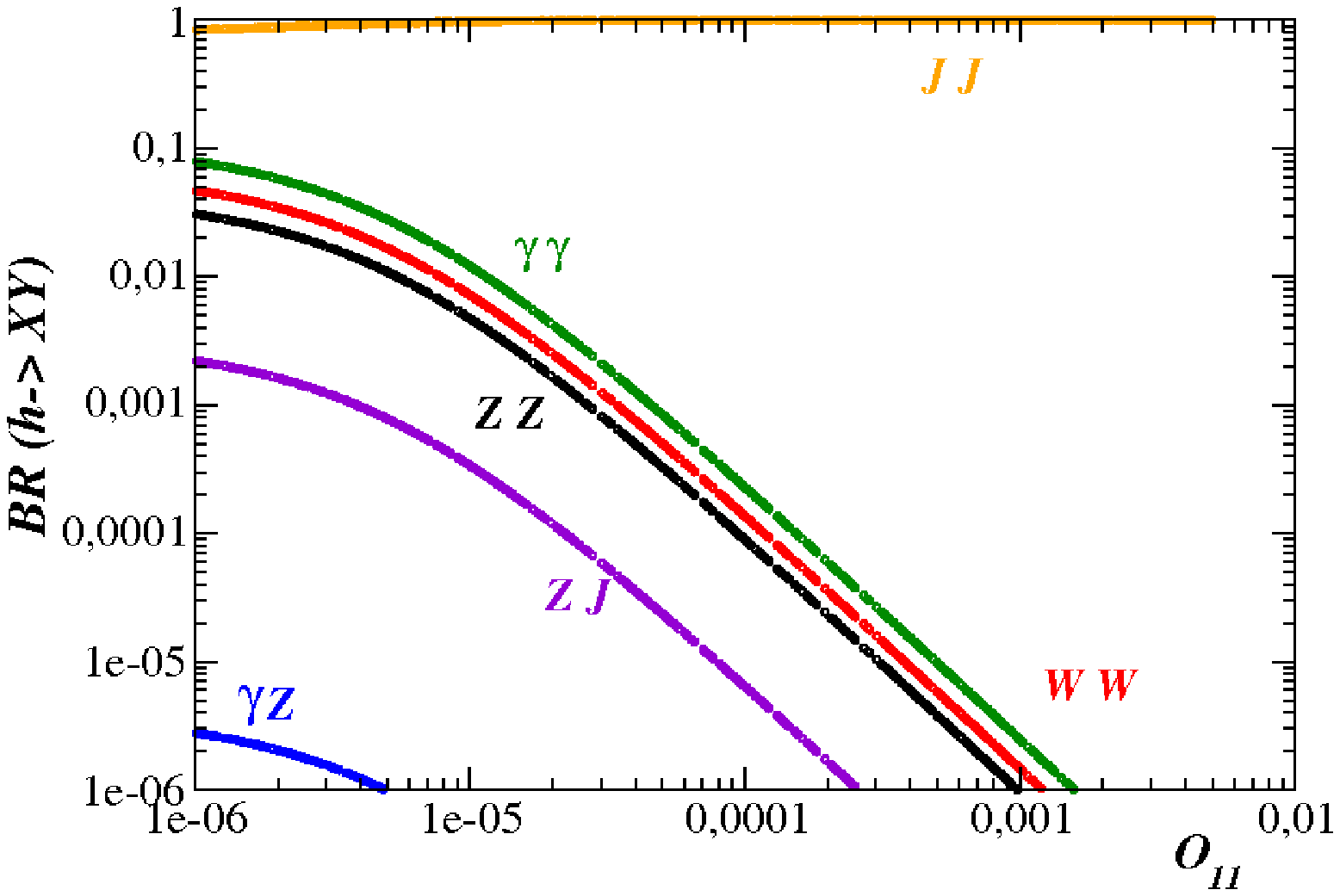}
\end{tabular}
\end{center}
\caption{\it Higgs  branching ratios  for scenario 2,  with $m_f=92.4$
  GeV.   Deviation from  exact  fermiophobia in  the  left frame,  and
  deviation from exact Majoron suppression in the right frame.
  \label{pheno:BR2}}
\end{figure}
%---------------------------------------------------------------------
%
In Fig.~\ref{pheno:BR2} we have scenario 2, also with a low Higgs mass
of $m_f=92.4$ GeV. In this case  the coupling $h_f JJ$ does not vanish
and  all  bosonic  decay   modes  decrease  monotonically,  while  the
fermionic         ones        increase.          In        particular,
$B(h_f\rightarrow\gamma\gamma)\approx 7\%$  for $O_R^{12}=0.0001$. The
behavior  of the  BR as  a function  of $O_R^{11}$  is similar  to the
previous case.
%
%------------------------ FIGURE -------------------------------------
\begin{figure}[!h]
\begin{center}
\begin{tabular}{cc}
\includegraphics[angle=0,width=8.1cm]{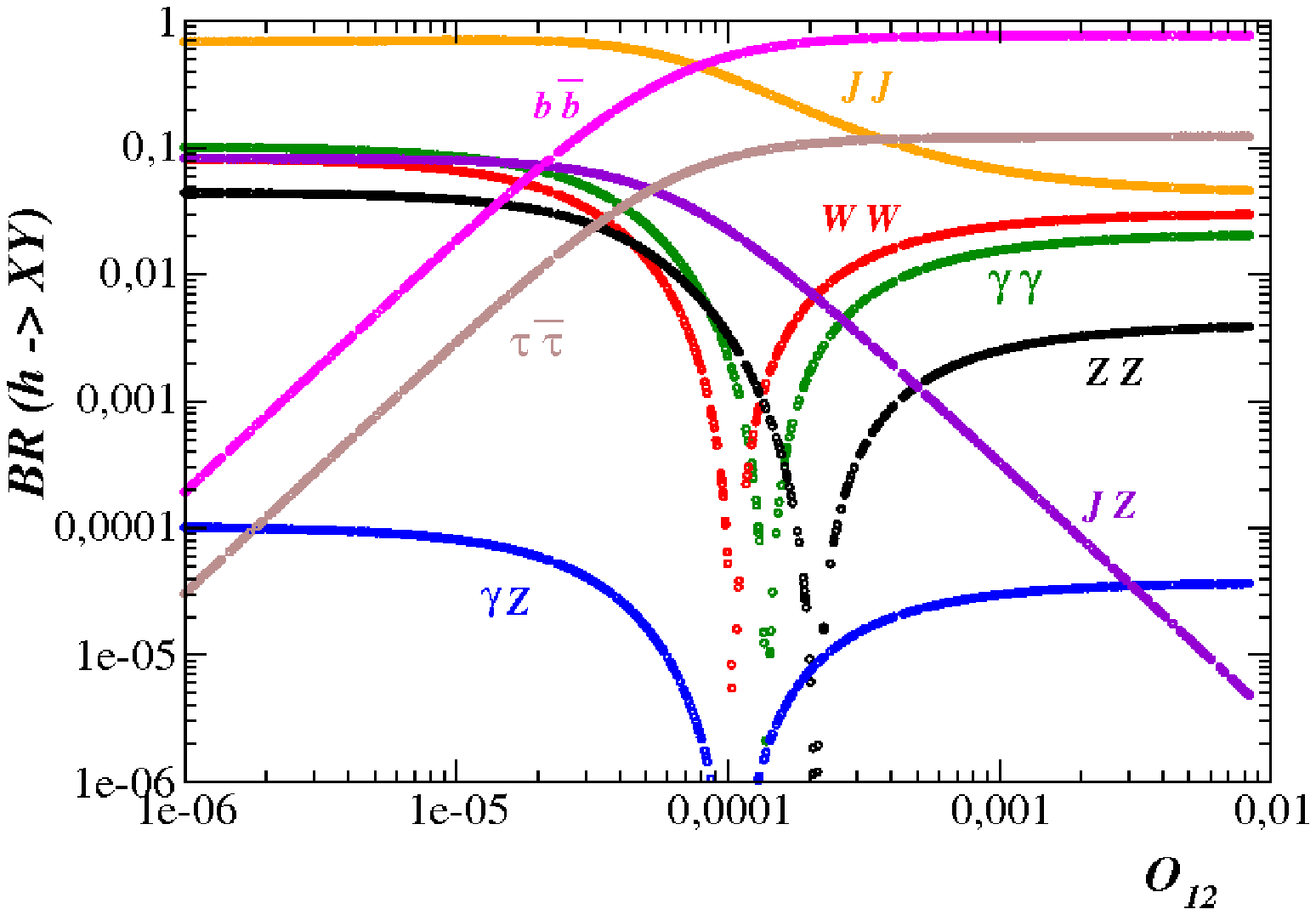}
&
\includegraphics[angle=0,width=8.3cm]{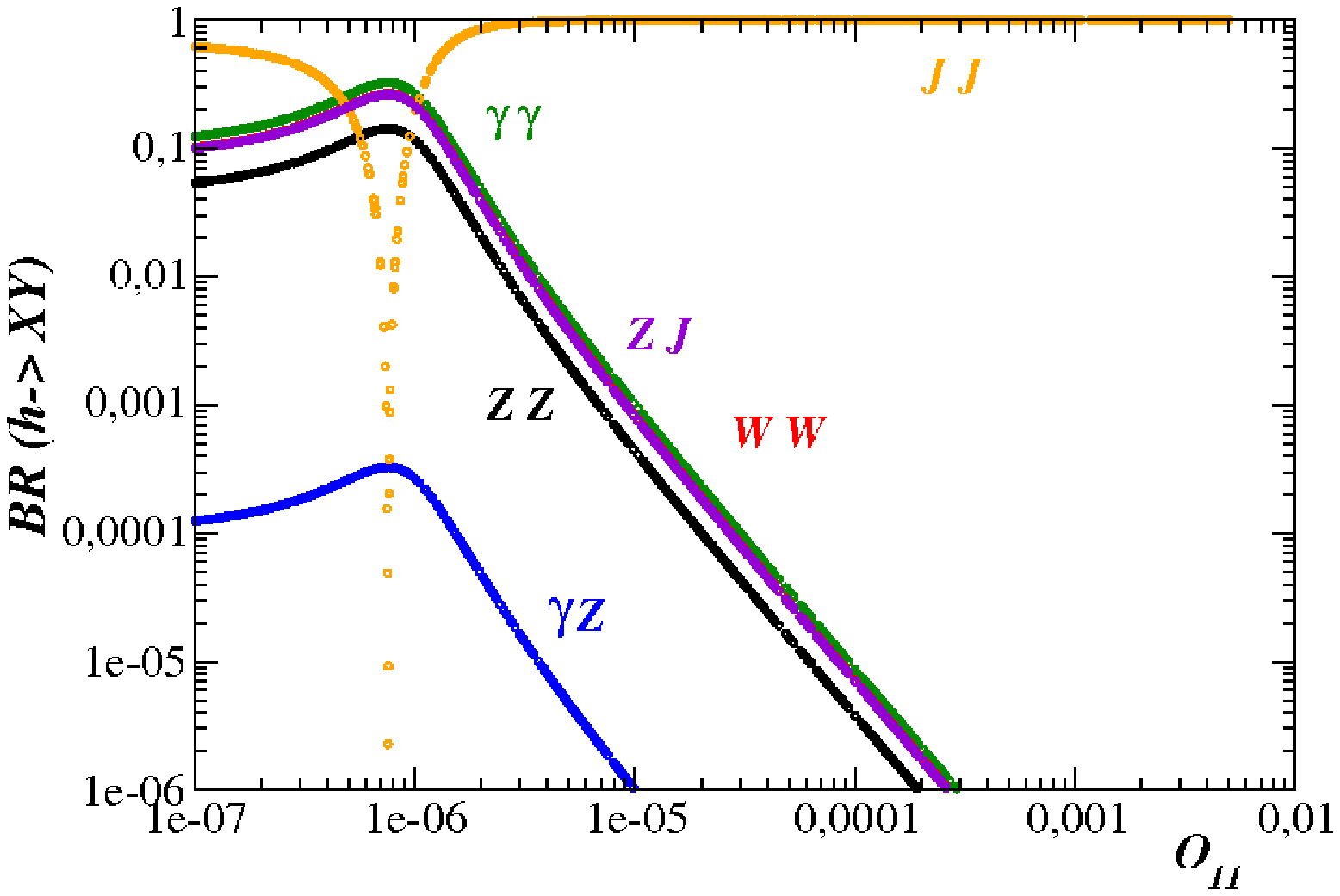}
\end{tabular}
\end{center}
\caption{\it Higgs  branching ratios for scenario  3, with $m_f=95.18$
  GeV.   Deviation from  exact  fermiophobia in  the  left frame,  and
  deviation from exact Majoron suppression in the right frame.
\label{pheno:BR3}}
\end{figure}
%---------------------------------------------------------------------
%
Scenario  3 is  analyzed in  Fig.~\ref{pheno:BR3}. This  is  the third
scenario with  a low Higgs  mass, $m_h=95.18$ GeV, which  enhances the
decay $h_f\rightarrow\gamma\gamma$. This  scenario is characterized by
the fact that all Higgs couplings  to a pair of gauge bosons vanish at
some point  in the displayed parameter  space. One effect  is that the
Higgs  decay  into  gauge  bosons  survives up  to  higher  values  of
$O_R^{12}$,  for  example  $h_f\rightarrow\gamma\gamma\approx2\%$  for
$O_R^{12}=0.01$, too small for the LHC, but useful for the ILC. In the
right  frame  we   see  that  the  coupling  $h_f   JJ$  vanishes  for
$O_R^{11}\approx 7\times 10^{-7}$, with  the effect that visible decay
modes  increase their  branching ratio.   After this  point  the decay
channels to the visible channels rapidly decrease.
%
%------------------------ FIGURE -------------------------------------
\begin{figure}[!h]
\begin{center}
\begin{tabular}{cc}
\includegraphics[angle=0,width=8.1cm]{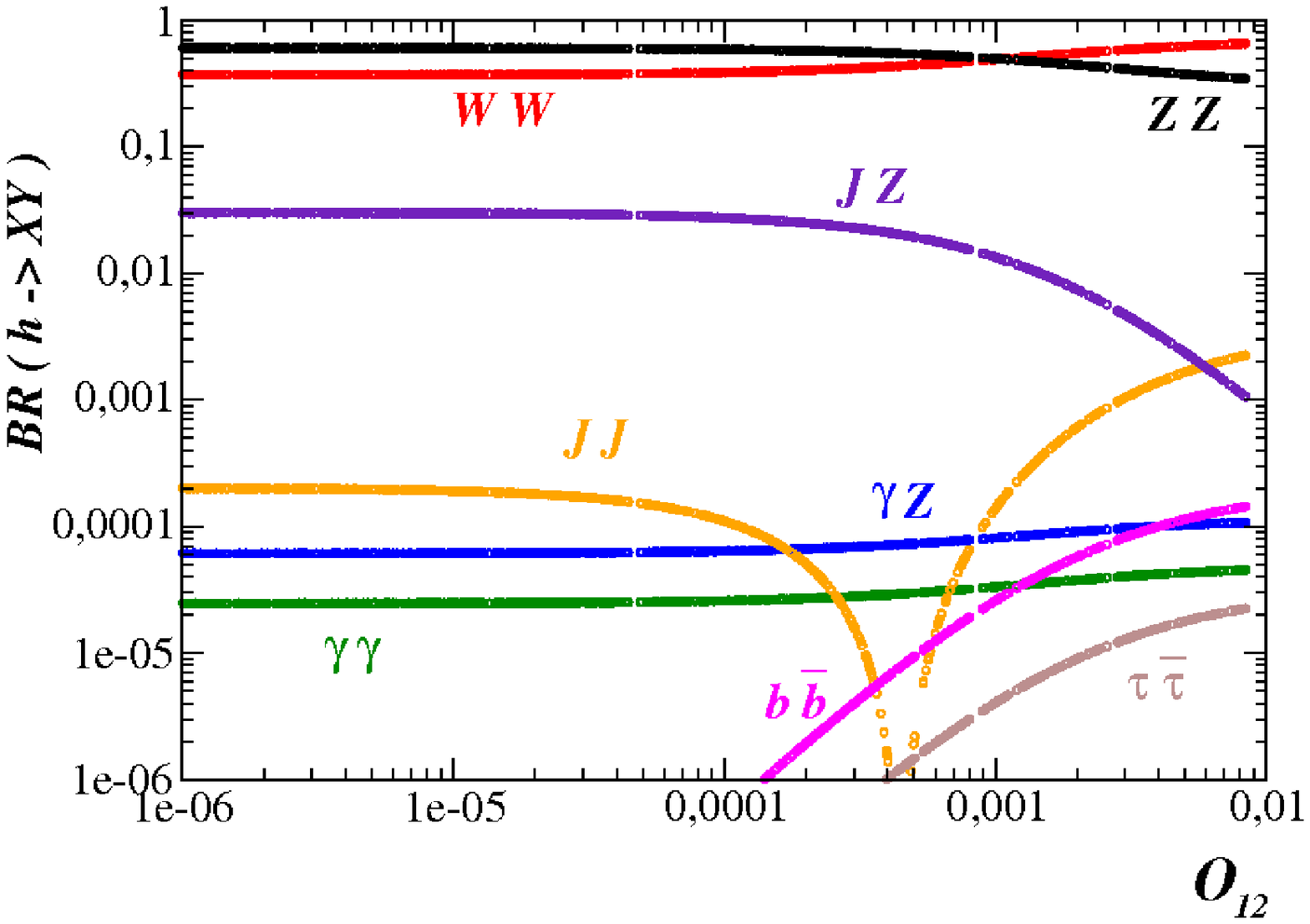}
&
\includegraphics[angle=0,width=8.3cm]{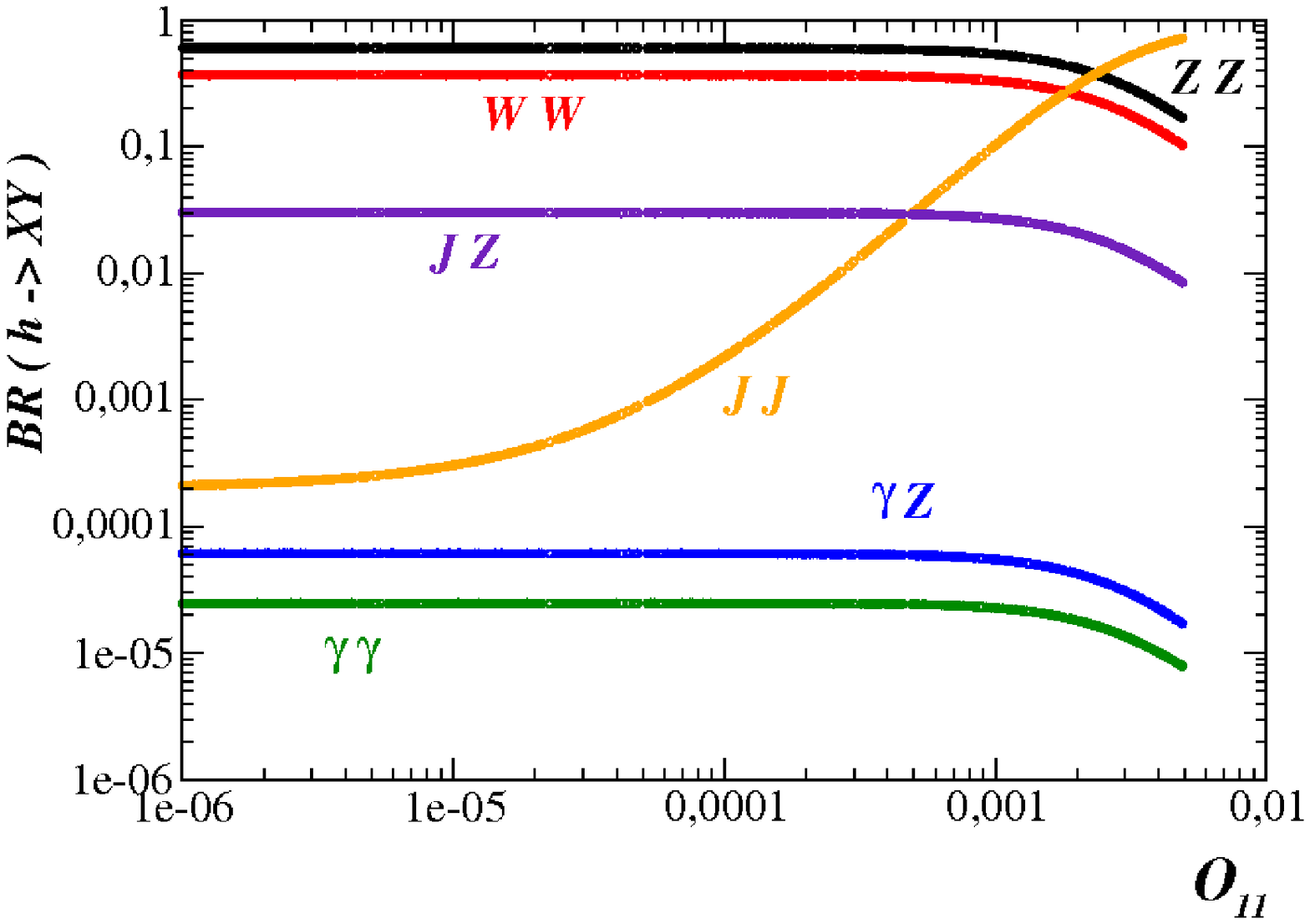}
\end{tabular}
\end{center}
\caption{\it Higgs branching ratios  for scenario 4, with $m_f=244.78$
  GeV.   Deviation from  exact  fermiophobia in  the  left frame,  and
  deviation from exact Majoron suppression in the right frame.
 \label{pheno:BR4}}
\end{figure}
%---------------------------------------------------------------------
%
In  the  following   two  figures  we  analyze  scenarios   4  and  5,
characterized  by   a  large  Higgs   mass,  where  the   decay  modes
$h_f\rightarrow  ZZ$ and  $h_f\rightarrow WW$  are very  important. In
Fig.~\ref{pheno:BR4} we have scenario 4, characterized by $m_f=244.78$
GeV. In the left frame we see that the decay modes $h_f\rightarrow ZZ$
and  $h_f\rightarrow WW$  remain  between $30\%$  and  $70\%$ when  we
deviate from  fermiophobic limit, even up to  $O_R^{12}\lsim 0.01$. In
addition, we  see that $B(h_f\rightarrow  ZZ)>B(h_f\rightarrow WW)$ up
to $O_R^{12}\approx0.001$,  while for larger values  of $O_R^{12}$ the
ratio returns  to the SM  one. In the  right frame, one see  that the
deviation from  exact Majoron suppression  causes the BR of  the decay
mode $h_f\rightarrow JJ$ to increase  until it becomes the largest one
for  $O_R^{11}\gsim  0.002$.    However,  $B(h_f\rightarrow  ZZ)$  and
$B(h_f\rightarrow  WW)$  never  go   below  $10\%$  in  the  displayed
parameter space.

%
%------------------------ FIGURE -------------------------------------
\begin{figure}[!h]
\begin{center}
\begin{tabular}{cc}
\includegraphics[angle=0,width=8.1cm]{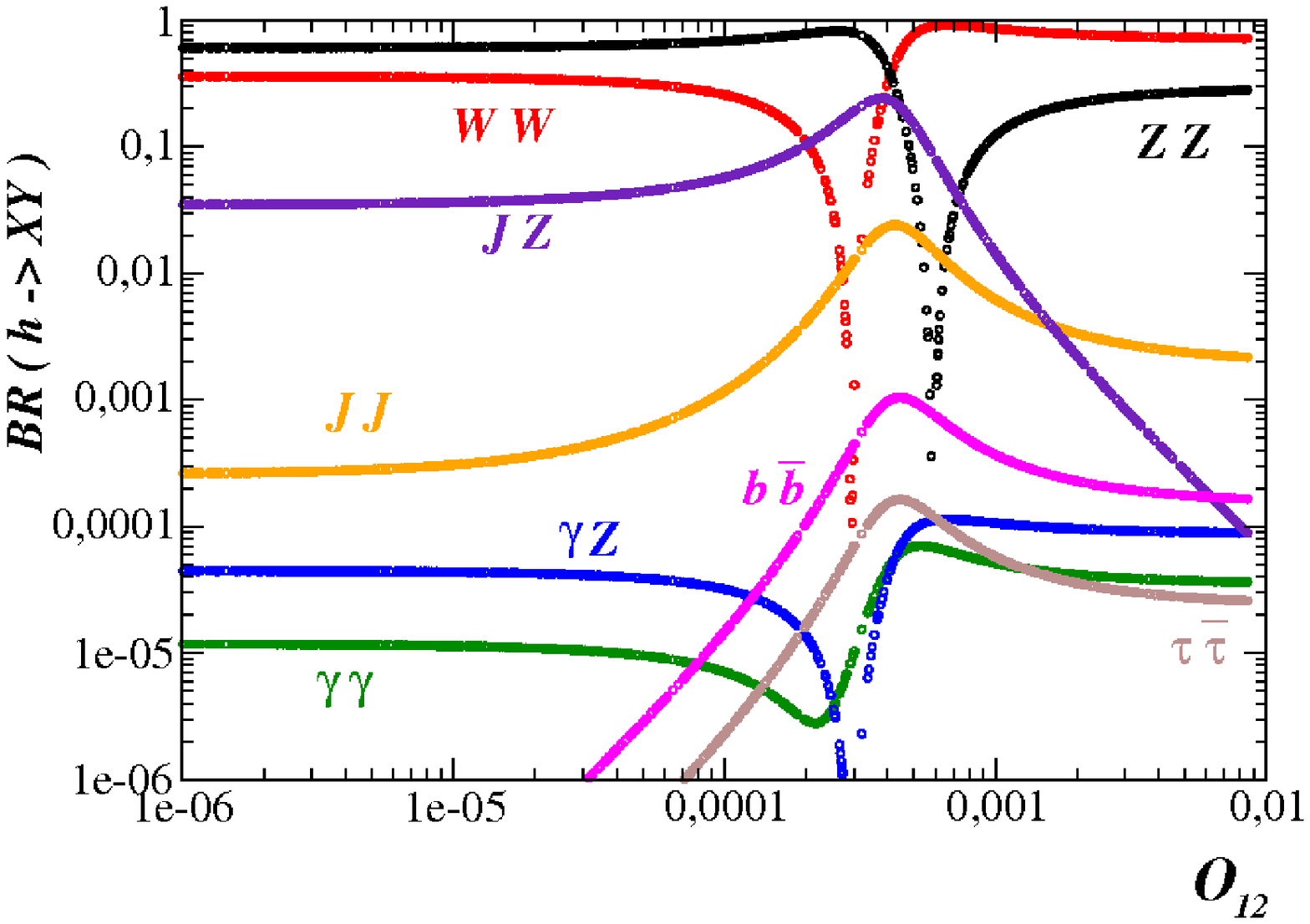}
&
\includegraphics[angle=0,width=8.3cm]{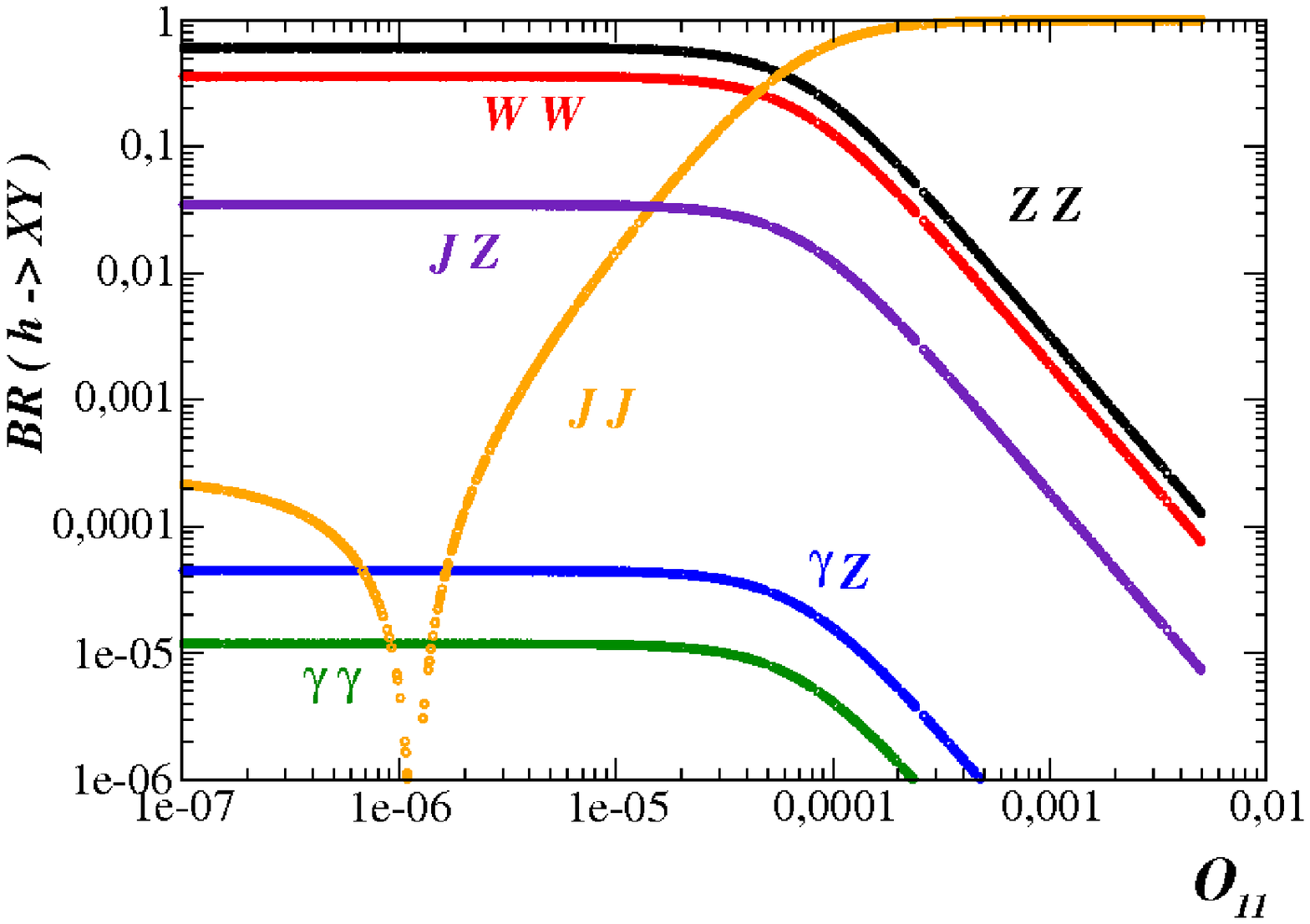}
\end{tabular}
\end{center}
\caption{\it Higgs branching ratios  for scenario 5, with $m_f=272.33$
  GeV.   Deviation from  exact  fermiophobia in  the  left frame,  and
  deviation from exact Majoron suppression in the right frame.
 \label{pheno:BR5}}
\end{figure}
%---------------------------------------------------------------------
%
The last  scenario 5, characterized  by $m_f=272.33$ GeV, is  shown in
Fig.~\ref{pheno:BR5}. We see in the left frame that the deviation from
exact  fermiophobia causes  the couplings  $h_f  WW$ and  $h_f ZZ$  to
vanish   for   $O_R^{12}\approx0.0003$  and   $O_R^{12}\approx0.0006$,
respectively.   The invisible decay  peaks up  to $2\%$  between these
zeros,  but  both $B(h_f\rightarrow  ZZ)$  and $B(h_f\rightarrow  WW)$
never fall  simultaneously below  $~30\%$. In the  right frame  we see
that $B(h_f\rightarrow  ZZ)$ and $B(h_f\rightarrow  WW)$ remain fairly
stable at $60\%$ and $30\%$  respectively when one deviates from exact
Majoron  suppression,   until  $h_f\rightarrow  JJ$   becomes  equally
important at $O_R^{11}\approx 5\times 10^{-5}$. After that point, $WW$
and $ZZ$ modes fall fast, reaching $5\%$ at $O_R^{11}\approx0.0002$.

We remind the reader that Table \ref{tab:O12} shows that $30\%$ of the
allowed  parameter   space  satisfies  fermiophobia  to   a  level  of
$|O_R^{12}|<0.0014$,  and  that  Table  \ref{tab:O13}  indicates  that
$36.6\%$   of  parameter  space   satisfy  fermiophobia   and  Majoron
suppression to a level  of $|O_R^{13}|>0.99999$. These numbers tell us
that  the model ``prefers''  fermiophobia and/or  Majoron suppression,
which in turn comes from the hierarchy of the three vacuum expectation
values.
%
%%%%%%%%%%%%%%%%%%%%%%%%%%%%%%%%%%%%%%%%%%%%%%%%%%%%%%%%%%%%%%%%%%%%%%
\section{Conclusions}
\label{sec:concl}
%%%%%%%%%%%%%%%%%%%%%%%%%%%%%%%%%%%%%%%%%%%%%%%%%%%%%%%%%%%%%%%%%%%%%%
%

We have  studied the  Higgs sector  of a model  with scalar  fields in
singlet, doublet  and triplet  isospin representations. In  this model
the singlet  scalar field acquires  a vacuum expectation  value, which
spontaneously breaks  lepton number and leads to  a vacuum expectation
value for  the scalar triplet field,  the latter providing  a mass for
the neutrinos.   The lightest  neutral CP-even scalar  ($H$) can  be a
fermiophobic  Higgs  boson  (i.e.    a  scalar  with  very  suppressed
couplings  to  fermions).   We  have studied  distinctive  signals  of
fermiophobia which  can be probed at  the LHC, and we  have shown that
fermiophobia  is  not a  fine-tuned  situation,  unlike  in Two  Higgs
Doublet  Models.  Characteristic  signals  are possible  for both  the
cases of  small and heavy Higgs  boson mass. For  a light fermiophobic
Higgs   boson    a   distinctive   signal   would    be   a   moderate
$B(H\rightarrow\gamma\gamma)$  accompanied by a  large $B(H\rightarrow
JJ)$ (where  $J$ is a Majoron),  the latter being  an invisible decay.
For the case of a large Higgs boson mass the decay modes $H\rightarrow
ZZ,  WW$ can  be  dominant,  while the  channel  $H\rightarrow JJ$  is
suppressed.  In  this situation,  $B(H\rightarrow ZZ)$ is  larger than
$B(H\rightarrow  WW)$,  which  differs  from  the  SM  prediction  and
provides a test for the model.

%%%%%%%%%%%%%%%%%%%%%%%%%%%%%%%%%%%%%%%%%%%%%%%%%%%%%%%%%%%%%%%%%%%%%%
\appendix
\section{}
\label{app:A}
%%%%%%%%%%%%%%%%%%%%%%%%%%%%%%%%%%%%%%%%%%%%%%%%%%%%%%%%%%%%%%%%%%%%%%

Here we present the relevant  Feynman rules which involve the lightest
Higgs boson  in our HTM.  The notation  used here is the  same as that
used in section \ref{sec:HTM}.

\vspace{0.3cm}
{\it Higgs-Higgs-Higgs Interactions}\\
%------------------------ FIGURE -------------------------------------
\begin{tabular}{llll}
\begin{picture}(150,70)(0,-10)
%left horizontal line
\DashLine(10,0)(60,0){5}
\Text(0,0)[c]{$H_i^0$}
%right horizontal line
\DashLine(60,0)(110,0){5}
\Text(120,0)[c]{$J$}
%upper vertical line
\DashLine(60,50)(60,0){5}
\Text(65,45)[l]{$J$}
%blob
\Vertex(60,0){2}
\end{picture}
&
\raisebox{30\unitlength}{
\begin{minipage}{5cm}
\lefteqn{
i \left( O^{i1}_{R} \left[ (O^{11}_{I})^2 v_1\beta_1 +
\half(O^{12}_{I})^2(v_3\kappa+v_1\beta_2) +
\half(O^{13}_{I})^2v_1\beta_3 +  \right. \right.
}
\lefteqn{
\left. O^{12}_{I}O^{13}_{I}v_2\kappa \right] +
 O^{i2}_{R} \left[ \half(O^{11}_{I})^2 v_2 \beta_2 
+ \half (O^{13}_{I})^2 v_2 (\lambda_5+\lambda_3)+ \right.
}
\lefteqn{
\left. (O^{12}_{I})^2 v_2 \lambda_1 + 
\kappa (v_1O^{12}_{I} O^{13}_{I}+v_2O^{11}_{I} O^{13}_{I}
+v_3O^{11}_{I} O^{12}_{I}) \right] +
}
\lefteqn{
 O^{i3}_{R}\left[ \half(O^{11}_{I})^2v_3\beta_3 +
 \half(O^{12}_{I})^2(v_1\kappa + v_3 (\lambda_5+\lambda_3)) +
 v_2\kappa O^{11}_{I}O^{12}_{I} +\right.
}
\lefteqn{
\left. \left.
(O^{13}_{I})^2 v_3(\lambda_2+\lambda_4)  \right]
\right)
}
\end{minipage}
}
\end{tabular}

\vspace{0.7cm}

% H0 H+ H-
%------------------------ FIGURE --------------------------------------------
\begin{tabular}{ll}
\begin{picture}(150,70)(0,-10)
%left horizontal line
\DashLine(10,0)(60,0){5}
\Text(0,0)[c]{$H_i^0$}
%right horizontal line
\DashArrowLine(110,0)(60,0){5}
\Text(120,0)[c]{$H^{-}$}
%upper vertical line
\DashArrowLine(60,50)(60,0){5}
\Text(65,45)[l]{$H^{+}$}
%blob
\Vertex(60,0){2}
\end{picture}
&
\raisebox{30\unitlength}{
\begin{minipage}{5cm}
\lefteqn{
i \left(
O^{i1}_{R} \left( s_+^2\beta_3v_1-\sqrt{2}s_+c_+\kappa v_2+c_+^2\beta_2 v_1 \right)
+\right.}
\lefteqn{
O^{i2}_{R} \left(
s_+^2(\lambda_3\!+\!\half\lambda_5)v_2\!-\! \half\sqrt{2}s_+c_+(2\kappa v_1\!-\!
\lambda_5v_3)\!+\!2c_+^2\lambda_1v_2\! \right)
+}
\lefteqn{
\left. O^{i3}_{R}\left(
2s_+^2(\lambda_2+\lambda_4)v_3+\half\sqrt{2}s_+c_+\lambda_5v_2+c_+^2\lambda_3v_3
\right)\right)
}
\end{minipage}
}
\end{tabular}
%-----------------------------------------------------------------------------

\vspace{0.3cm}
% H0 H++ H--
%------------------------ FIGURE --------------------------------------------
\begin{tabular}{ll}
\begin{picture}(150,70)(0,-10)
%left horizontal line
\DashLine(10,0)(60,0){5}
\Text(0,0)[c]{$H_i^0$}
%right horizontal line
\DashArrowLine(110,0)(60,0){5}
\Text(120,0)[c]{$\Delta^{--}$}
%upper vertical line
\DashArrowLine(60,50)(60,0){5}
\Text(65,45)[l]{$\Delta^{++}$}
%blob
\Vertex(60,0){2}
\end{picture}
&
\raisebox{35\unitlength}{
\begin{minipage}{5cm}
\lefteqn{
i\left(O^{i1}_{R}\beta_3v_1 + O^{i2}_{R} \lambda_3v_2 
+ 2O^{i3}_{R}\lambda_2v_3\right)
}
\end{minipage}
}
\end{tabular}
%-----------------------------------------------------------------------------

\vspace{1cm}

{\it Higgs-gauge boson-Higgs Interactions}\\

% Z H0 J
%------------------------ FIGURE --------------------------------------------
\begin{tabular}{ll}
\begin{picture}(150,70)(0,-10)
%left horizontal line
\Photon(10,0)(60,0){3}{4}
\Text(0,0)[c]{$Z_\mu^0$}
%right horizontal line
\DashArrowLine(110,0)(60,0){5}
\Text(120,0)[c]{$H_i^0$}
\Text(90,10)[c]{$k$}
%upper vertical line
\DashArrowLine(60,50)(60,0){5}
\Text(65,45)[l]{$J$}
\Text(55,30)[r]{$p$}
%blob
\Vertex(60,0){2}
\end{picture}
&
\raisebox{35\unitlength}{
\begin{minipage}{5cm}
\lefteqn{
ic\,v_2\,v_3 \sqrt{g^2+g'^2}(v_3O^{i2}_{R}-v_2 O^{i3}_{R})
(p-k)^\mu
}
\end{minipage}
}
\end{tabular}
%-----------------------------------------------------------------------------

% Z H0 A
%------------------------ FIGURE --------------------------------------------
\begin{tabular}{ll}
\begin{picture}(150,70)(0,-10)
%left horizontal line
\Photon(10,0)(60,0){3}{4}
\Text(0,0)[c]{$Z_\mu^0$}
%right horizontal line
\DashArrowLine(110,0)(60,0){5}
\Text(120,0)[c]{$H_i^0$}
\Text(90,10)[c]{$k$}
%upper vertical line
\DashArrowLine(60,50)(60,0){5}
\Text(65,45)[l]{$A$}
\Text(55,30)[r]{$p$}
%blob
\Vertex(60,0){2}
\end{picture}
&
\raisebox{35\unitlength}{
\begin{minipage}{5cm}
\lefteqn{
i\frac{b}{2v_3}\sqrt{g^2+g'^2}(v_3O^{i2}_{R}-v_2 O^{i3}_{R})
(p-k)^\mu
}
\end{minipage}
}
\end{tabular}
%-----------------------------------------------------------------------------

% Z H+ H-
%------------------------ FIGURE --------------------------------------------
\begin{tabular}{ll}
\begin{picture}(150,70)(0,-10)
%left horizontal line
\Photon(10,0)(60,0){3}{4}
\Text(0,0)[c]{$Z_\mu^0$}
%right horizontal line
\DashArrowLine(110,0)(60,0){5}
\Text(120,0)[c]{$H^-$}
\Text(90,10)[c]{$k$}
%upper vertical line
\DashArrowLine(60,50)(60,0){5}
\Text(65,45)[l]{$H^+$}
\Text(55,30)[r]{$p$}
%blob
\Vertex(60,0){2}
\end{picture}
&
\raisebox{35\unitlength}{
\begin{minipage}{5cm}
\lefteqn{
i\frac{g'^2(v_2^2+v_3^2) - g^2v_3^2}{\sqrt{g^2+g'^2}(2v_3^2+v_2^2)}
(p-k)^\mu
}
\end{minipage}
}
\end{tabular}
%-----------------------------------------------------------------------------

% Z H++ H--
%------------------------ FIGURE --------------------------------------------
\begin{tabular}{ll}
\begin{picture}(150,70)(0,-10)
%left horizontal line
\Photon(10,0)(60,0){3}{4}
\Text(0,0)[c]{$Z_\mu^0$}
%right horizontal line
\DashArrowLine(110,0)(60,0){5}
\Text(120,0)[c]{$\Delta^{--}$}
\Text(90,10)[c]{$k$}
%upper vertical line
\DashArrowLine(60,50)(60,0){5}
\Text(65,45)[l]{$\Delta^{++}$}
\Text(55,30)[r]{$p$}
%blob
\Vertex(60,0){2}
\end{picture}
&
\raisebox{35\unitlength}{
\begin{minipage}{5cm}
\lefteqn{
-i\frac{g^2 - g'^2}{\sqrt{g^2+g'^2}}
(p-k)^\mu
}
\end{minipage}
}
\end{tabular}
%-----------------------------------------------------------------------------

% Y H+ H-
%------------------------ FIGURE --------------------------------------------
\begin{tabular}{ll}
\begin{picture}(150,70)(0,-10)
%left horizontal line
\Photon(10,0)(60,0){3}{4}
\Text(0,0)[c]{$\gamma_\mu$}
%right horizontal line
\DashArrowLine(110,0)(60,0){5}
\Text(120,0)[c]{$H^-$}
\Text(90,10)[c]{$k$}
%upper vertical line
\DashArrowLine(60,50)(60,0){5}
\Text(65,45)[l]{$H^+$}
\Text(55,30)[r]{$p$}
%blob
\Vertex(60,0){2}
\end{picture}
&
\raisebox{35\unitlength}{
\begin{minipage}{5cm}
\lefteqn{
-\,i\,e\,(p-k)^\mu
}
\end{minipage}
}
\end{tabular}
%-----------------------------------------------------------------------------

% Y H++ H--
%------------------------ FIGURE --------------------------------------------
\begin{tabular}{ll}
\begin{picture}(150,70)(0,-10)
%left horizontal line
\Photon(10,0)(60,0){3}{4}
\Text(0,0)[c]{$\gamma_\mu$}
%right horizontal line
\DashArrowLine(110,0)(60,0){5}
\Text(120,0)[c]{$\Delta^{--}$}
\Text(90,10)[c]{$k$}
%upper vertical line
\DashArrowLine(60,50)(60,0){5}
\Text(65,45)[l]{$\Delta^{++}$}
\Text(55,30)[r]{$p$}
%blob
\Vertex(60,0){2}
\end{picture}
&
\raisebox{35\unitlength}{
\begin{minipage}{5cm}
\lefteqn{
-2\,i\,e\,(p-k)^\mu
}
\end{minipage}
}
\end{tabular}
%-----------------------------------------------------------------------------

\vspace{1cm}

{\it Higgs-gauge boson-gauge boson Interactions}\\

% Z Z H0
%------------------------ FIGURE --------------------------------------------
\begin{tabular}{ll}
\begin{picture}(150,70)(0,-10)
%left horizontal line
\Photon(10,0)(60,0){3}{4}
\Text(0,0)[c]{$Z_\mu^0$}
%right horizontal line
\Photon(110,0)(60,0){3}{4}
\Text(120,0)[c]{$Z_\nu^0$}
%upper vertical line
\DashArrowLine(60,50)(60,0){5}
\Text(65,45)[l]{$H_i^0$}
%blob
\Vertex(60,0){2}
\end{picture}
&
\raisebox{35\unitlength}{
\begin{minipage}{5cm}
\lefteqn{
i\frac{(g^2+g'^2)}{2}(O^{i2}_{R}v_2 + 4O^{i3}_{R}v_3)g^{\mu \nu}
}
\end{minipage}
}
\end{tabular}
%----------------------------------------------------------------------------

% W W H0
%------------------------ FIGURE --------------------------------------------
\begin{tabular}{ll}
\begin{picture}(150,70)(0,-10)
%left horizontal line
\Photon(10,0)(60,0){3}{4}
\Text(0,0)[c]{$W_\mu^+$}
%right horizontal line
\Photon(110,0)(60,0){3}{4}
\Text(120,0)[c]{$W_\nu^-$}
%upper vertical line
\DashArrowLine(60,50)(60,0){5}
\Text(65,45)[l]{$H_i^0$}
%blob
\Vertex(60,0){2}
\end{picture}
&
\raisebox{35\unitlength}{
\begin{minipage}{5cm}
\lefteqn{
i\frac{g^2}{2}(O^{i2}_{R}v_2 + 2O^{i3}_{R}v_3)g^{\mu \nu}
}
\end{minipage}
}
\end{tabular}
%----------------------------------------------------------------------------

\vspace{1cm}

{\it Higgs-Higgs- gauge boson-gauge boson Interactions}\\
\vspace{2cm}

% Y Y H+ H-
%------------------------ FIGURE --------------------------------------------
\begin{tabular}{llll}
\begin{picture}(150,70)(0,-10)
%left horizontal line
\Photon(10,70)(60,70){3}{4}
\Text(0,70)[c]{$\gamma_{\mu}$}
%right horizontal line
\Photon(60,70)(110,70){3}{4}
\Text(120,70)[c]{$\gamma_{\nu}$}
%upper vertical line
\DashArrowLine(60,120)(60,70){5}
\Text(65,115)[l]{$H^+$}
%Lower vertical line
\DashArrowLine(60,70)(60,20){5}
\Text(65,25)[l]{$H^-$}
%blob
\Vertex(60,70){2}
\end{picture}
&
\raisebox{75\unitlength}{
\begin{minipage}{5cm}
\lefteqn{
2\,i\,e^2\,g^{\mu \nu}
}
\end{minipage}
}
%&
\end{tabular}
%----------------------------------------------------------------------------

\vspace{1.2cm}

% Y Y H++ H--
%------------------------ FIGURE --------------------------------------------
\begin{tabular}{ll}
\begin{picture}(150,70)(0,-10)
%left horizontal line
\Photon(10,70)(60,70){3}{4}
\Text(0,70)[c]{$\gamma_{\mu}$}
%right horizontal line
\Photon(60,70)(110,70){3}{4}
\Text(120,70)[c]{$\gamma_{\nu}$}
%upper vertical line
\DashArrowLine(60,120)(60,70){5}
\Text(65,115)[l]{$\Delta^{++}$}
%Lower vertical line
\DashArrowLine(60,70)(60,20){5}
\Text(65,25)[l]{$\Delta^{--}$}
%blob
\Vertex(60,70){2}
\end{picture}
&
\raisebox{75\unitlength}{
\begin{minipage}{5cm}
\lefteqn{
8\,i\,e^2\,g^{\mu \nu}
}
\end{minipage}
}
\end{tabular}
%----------------------------------------------------------------------------

%%%%%%%%%%%%%%%%%%%%%%%%%%%%%%%%%%%%%%%%%%%%%%%%%%%%%%%%%%%%%%%%%%%%%%
\section*{Acknowledgments}  
%%%%%%%%%%%%%%%%%%%%%%%%%%%%%%%%%%%%%%%%%%%%%%%%%%%%%%%%%%%%%%%%%%%%%%
%

A.G.A.   was supported by  the ``National  Central University  plan to
develop  first-class universities'',  and  by a  Marie Curie  Incoming
International  Fellowship, FP7-PEOPLE-2009-IIF,  Contract  No. 252263.
M.A.D.  was  supported by Fondecyt  Regular Grant \#  1100837.  M.A.R.
was supported by  Fondecyt grant No.  3090069.  D.R.  was supported by
CONICYT.

%
%%%%%%%%%%%%%%%%%%%%%%%%%%%%%%%%%%%%%%%%%%%%%%%%%%%%%%%%%%%%%%%%%%%%%%

\end{document}